\documentclass{emulateapj}

\usepackage{natbib}
\usepackage{color}
\usepackage{natbib}
\usepackage{apjfonts}
\usepackage{xfrac}
\usepackage[dvipsnames]{xcolor}
\usepackage{ulem}
\usepackage{soul}
\usepackage{amsmath,xspace}
\usepackage{hyperref}
\hypersetup{linkcolor=red,citecolor=blue,filecolor=cyan,urlcolor=magenta}

\newcommand{\Mbh}{M_\mathrm{BH}}
\newcommand{\mbh}{$\Mbh$\xspace}
\newcommand{\hst}{{\it HST}\xspace}
\newcommand{\galfit}{{\sc Galfit}\xspace}

\shorttitle{Stellar Dynamics of NGC\,4151}
\shortauthors{Roberts, et al.}

\received{}
\accepted{}

\begin{document}

\title{The Black Hole Mass of NGC\,4151 from Stellar Dynamical Modeling}

\author{ Caroline~A.~Roberts\altaffilmark{1,2},
Misty~C.~Bentz\altaffilmark{1},
Eugene~Vasiliev\altaffilmark{3,4,5},
Monica~Valluri\altaffilmark{6},
and Christopher~A.~Onken\altaffilmark{7}
}

\altaffiltext{1}{Department of Physics and Astronomy,
		 Georgia State University,
		 Atlanta, GA 30303, USA;
		 roberts@astro.gsu.edu}
\altaffiltext{2}{Department of Physics and Astronomy,
                 University of Iowa,
                 Iowa City, IA, 52242, USA}
\altaffiltext{3}{Institute of Astronomy, 
        Madingley Road, Cambridge, CB3 0HA, UK}
\altaffiltext{4}{Rudolf Peierls Centre for Theoretical Physics, 
        1 Keble Road, Oxford, OX1 3NP, UK}
\altaffiltext{5}{Lebedev Physical Institute, 
        Leninsky Prospekt 53, Moscow, 119991, Russia}
\altaffiltext{6}{Department of Astronomy,
                 University of Michigan,
                 Ann Arbor, MI, 48109, USA}
\altaffiltext{7}{Research School of Astronomy and Astrophysics, 
        Australian National University, 
        Canberra, ACT 2611, Australia }

\begin{abstract}

The mass of a supermassive black hole (\mbh) is a fundamental property that can be obtained through observational methods. Constraining \mbh\ through multiple methods for an individual galaxy is important for verifying the accuracy of different techniques, and for investigating the assumptions inherent in each method. However, there exist only a few galaxies where multiple \mbh\ measurement techniques can be applied. NGC\,4151 is one of these rare galaxies for which multiple methods can be used: stellar and gas dynamical modeling because of its proximity ($D=15.8\pm0.4$\,Mpc from Cepheids), and reverberation mapping because of its active accretion. In this work, we re-analyzed $H-$band integral field spectroscopy of the nucleus of NGC\,4151 from  Gemini NIFS, improving the analysis at several key steps. 
We then constructed a wide range of axisymmetric dynamical models with the new orbit-superposition code \textsc{Forstand}. One of our primary goals is to quantify the systematic uncertainties in $\Mbh$ arising from different combinations of the deprojected density profile, inclination, intrinsic flattening, and mass-to-light ratio. As a consequence of uncertainties on the stellar luminosity profile arising from the presence of the AGN, our constraints on \mbh are rather weak. Models with a steep central cusp are consistent with no black hole; however, in models with more moderate cusps, the black hole mass lies within the range of  $0.25\times10^7\,M_\odot \lesssim \Mbh \lesssim 3\times10^7\,M_\odot$.
This measurement is somewhat smaller than the earlier analysis presented by \citet{2014ApJ...791...37O}, but agrees with previous \mbh values from gas dynamical modeling and reverberation mapping. Future dynamical modeling of reverberation data, as well as IFU observations with JWST, will aid in further constraining the mass of the central supermassive black hole in NGC\,4151.

\end{abstract}


\section{Introduction}
Two of the most convincing pieces of evidence for the existence of supermassive black holes are also based on two unique methods for measuring black hole mass (\mbh). The most robust black hole mass measurement is that based on proper motion studies of stars in the Galactic Center around Sagittarius A*, which require a supermassive black hole with $M_{\rm BH}=4 \times 10^{6}$\,M$_{\odot}$  \citep{ghez2000,genzel2000,2002Natur.419..694S,2019A&A...625L..10G}. Unfortunately, observing the proper motions of individual stars in the centers of other galaxies is not possible due to the distances involved. The second robust black hole mass measurement is in the case of M87 which is based on image reconstruction using very-long baseline interferometry that provided the first image of emission from just outside the event horizon. These observations allowed the black hole mass to be constrained through comparison with general relativistic magnetohydrodynamic models \citep{2019ApJ...875L...1E}. Unfortunately, imaging of the event horizon is only currently possible for M87 and for Sagittarius A*, which has a comparable angular extent on the sky. All other black holes in the nearby universe are too small and/or too distant to be resolved with current technology, and so other techniques must be used to study additional supermassive black holes. 

Another direct method for constraining \mbh is the use of water maser emission in edge-on nuclear gas disks  \citep{1995Natur.373..127M,2005ApJ...629..719H,2016ApJ...826L..32G}. The maser emission in the disk traces out the Keplerian rotation curve of the gas, constraining the enclosed mass. Unfortunately, water maser emission is quite rare and also requires a specific set of circumstances to be fulfilled before it can be used for \mbh measurements, so there are only a few galaxies where this technique may be used.  

For larger samples of black hole masses, there are three other established direct \mbh measurement techniques. Gas dynamical (GD) modeling relies on spatially-resolved gas kinematics in galactic nuclei to infer the geometry and inclination of the gas as well as the enclosed mass (e.g., \citealt{1997ApJ...489..579M,2015ApJ...809..101D}), though the gas may be affected by non-circular motions and turbulence. Stellar dynamical (SD) modeling is similar, but considers a more general kinematic structure of the stellar motion, not limited to circular orbits (e.g.,  \citealt{1998ApJ...493..613V,1999ApJ...514..704C,2003ApJ...583...92G,2004ApJ...602...66V}). Reverberation mapping (RM, \citealt{1982ApJ...255..419B,1993PASP..105..247P}) uses light echoes in the emission from active galactic nuclei (AGN)  to constrain the kinematics of gas on spatially-unresolvable scales deep in the nuclear region.  Both GD and SD modeling require high spatial resolution and are dependent on the distance to the galaxy. RM instead requires high temporal sampling and is distance independent. 

For active galaxies, RM is the most prevalent \mbh measurement technique.  Continuum emission, likely arising from the accretion disk, travels outwards at the speed of light and photoionizes gas in the broad line region (BLR), where it is processed and re-emitted as spectral lines. Variability in the continuum flux (most likely from instabilities in the disk and/or variable accretion rates) thus causes variability in the broad line flux as well. Through spectrophotometric monitoring, the time delay $\tau$ between variations in the continuum radiation and the response in the broad emission lines can be measured.  The recombination timescale of the BLR is very short compared to typical time delays, so $\tau$ is simply the light-crossing time and $c \tau$ is the  responsivity-weighted mean radius of the BLR. When  combined with a constraint on the Doppler velocities of the BLR gas ($V$), \mbh can be determined via the virial theorem:
\begin{equation}
M_{\rm BH} = f\frac{c\tau V^2}{G},
\end{equation}
where $f$ is an order-unity scaling factor that depends on the detailed geometry and kinematics of the BLR gas and $G$ is the gravitational constant. RM is a direct method of determining \mbh based on the gravitational influence of the black hole on a luminous tracer, but its most common application relies on an average $f$ factor that is derived from comparison with dynamical modeling (e.g., \citealt{2004ApJ...615..645O,2013ApJ...773...90G,2017ApJ...838L..10B}).

RM masses determined in this way rely on certain assumptions, such as a symmetric geometry of the BLR and the assumption that gravity dominates the kinematics of the BLR gas. The BLR exhibits ionization stratification, in that high ionization lines are observed to have shorter time delays than low ionization lines. These short time delays are accompanied by high velocities in the line widths.  For those cases when it has been possible to explore the relationship between $\tau$ and $V$ for multiple broad emission lines in the same AGN, the measurements have been consistent with $V \propto \tau^{-0.5}$, as expected for a virial relationship (e.g., \citealt{peterson2004,kollatschny2003,2010ApJ...716..993B}) and thus supporting the assumption that gravity dominates the kinematics of the region.  
Velocity-resolved RM \citep{2014MNRAS.445.3073P,2017ApJ...849..146G}, on the other hand, allows the full geometry and kinematics of the BLR to be constrained and avoids many of the assumptions involved in  using a mean time delay and adopting a typical $f$ factor, thus providing a direct, primary constraint on \mbh. 

SD modeling is generally applied to quiescent galaxies -- elliptical or spheroidal galaxies and the bulges of disk galaxies.  It is a direct, primary method that constrains \mbh by fitting the bulk kinematics of stars derived from spatially-resolved spectroscopy with simulated  kinematics constructed for a galaxy model with a similar surface brightness density profile to the observed galaxy. 
Several different approaches have been employed for stellar dynamical modeling, e.g., the solution of the spherical or axisymmetric Jeans equation \citep{vandermarel_1994,cappellari_2008, cappellari_2014, cappellari_2020}, distribution function fitting \citep{vandermarel_1994,Magorrian_2019}, guided (``made-to-measure'') $N$-body simulations \citep{1996MNRAS.282..223S,2013MNRAS.429.2974D}, and the Schwarzschild orbit-superposition method \citep{1979ApJ...232..236S,1993ApJ...409..563S}.
In the past two decades, in conjunction with spatially-resolved absorption line spectroscopy, increasingly sophisticated versions of the orbit-superposition method \citep{1998ApJ...493..613V, 1999ApJ...514..704C, gebhardt_etal_2003, 2004ApJ...602...66V,
2004MNRAS.353..391T, 2008MNRAS.385..647V, 2010MNRAS.401.1770V, 2020ApJ...889...39V, 2021MNRAS.500.1437N} have been used to construct self-consistent dynamical models of galactic nuclei and to derive their black hole masses, stellar mass-to-light ratios $\Upsilon$, and internal orbit distributions. Consequently, orbit superposition is now the most widely used method for black hole mass determination and is responsible for obtaining the majority of BH mass measurements from dynamical modeling to date \citep[e.g.][]{2013ApJ...764..184M,2016ApJ...818...47S}. 
The accuracy of any SD method depends on the ability to spatially resolve the SMBH sphere of influence (the region where the gravity of the SMBH dominates over the gravity of the stars), 
although some authors argue that resolving the sphere of influence is not strictly necessary (e.g., see \citealt{2006ApJ...646..754D,Gultekin_2011}). 

Like SD modeling, RM is also a prolific measurement technique, and has been used to determine masses of $\sim$100 supermassive black holes (e.g., \citealt{2015PASP..127...67B,grier2017}). However, for the vast majority of targets it is not possible to constrain \mbh through both RM and SD modeling: bright broad-lined AGNs are rare in the local universe, so they are generally too far away to achieve the spatial resolution needed for dynamical modeling.  This is an important point, because tens of thousands of indirect measurements have been made by large surveys and are used to explore the growth and evolution of black holes and galaxies throughout cosmic history based on the information gleaned from these smaller samples of direct \mbh measurements, {\it yet we do not currently know if RM and SD modeling provide consistent mass measurements when applied to the same galaxies.} 

Direct comparisons of RM and SD modeling through \mbh measurements of the same targets, with their different assumptions, biases, and data and technical requirements, have only been accomplished for two galaxies to date: NGC\,4151 \citep{2006ApJ...651..775B,2014ApJ...791...37O,2018ApJ...866..133D} and NGC\,3227 \citep{2006ApJ...646..754D,2010ApJ...721..715D}.  We have thus undertaken an effort to improve and increase the sample of high-quality \mbh determinations for the small sample of nearby, bright Seyfert 1 galaxies where RM and SD modeling may be directly compared.
 
An important step in this program is a complete reanalysis of the SD modeling results for NGC\,4151, which we describe in this manuscript.  NGC\,4151 is the brightest Seyfert~1 in the northern hemisphere and one of the nearest AGNs at $z=0.0033$. NGC\,4151 has been studied with both SD modeling \citep{2007ApJ...670..105O,2014ApJ...791...37O} and RM \citep{2006ApJ...651..775B,2018ApJ...866..133D}, as well as GD modeling (using H$_{2}$; \citealt{2008ApJS..174...31H}). The sphere of influence of NGC\,4151 is estimated to be $\sim$15\, parsecs, or $\sim0\farcs2$, which is well-matched to the spatial resolution that may be achieved from the ground with adaptive optics.

The original SD modeling analysis by \citet{2014ApJ...791...37O} did not take full advantage of the spatial resolution available with the instrumentation, and we have identified several additional improvements in the data reduction and analysis that we have implemented and describe below. On the modeling side, we use a powerful new orbit-superposition code \textsc{Forstand} \citep{2020ApJ...889...39V} that overcomes many limitations of other stellar-dynamical modeling codes (e.g. restriction to axisymmetry, not accounting for dark matter and inability to account for figure rotation, limited size of orbit libraries, etc.), although in the present study we do not use all its novel features. 

In Section~\ref{sec:obs} of this paper, we introduce relevant integral field spectroscopy and imaging observations. Section~\ref{sec:phot} details the analysis of the photometry while Section~\ref{sec:kin} details the analysis of the two spectroscopic data sets. In Section~\ref{sec:models} we describe the application of the \textsc{Forstand} modeling code to NGC\,4151. Section~\ref{sec:diss} compares the modeling results and conclusions for \mbh with previous studies of NGC\,4151, discusses the obstacles of stellar dynamical modeling in this application, and shares the upcoming direction of this project and related research. We summarize in Section~\ref{sec:sum}.

\section{Observations} \label{sec:obs}

\begin{figure}
	\includegraphics[width=3.3in]{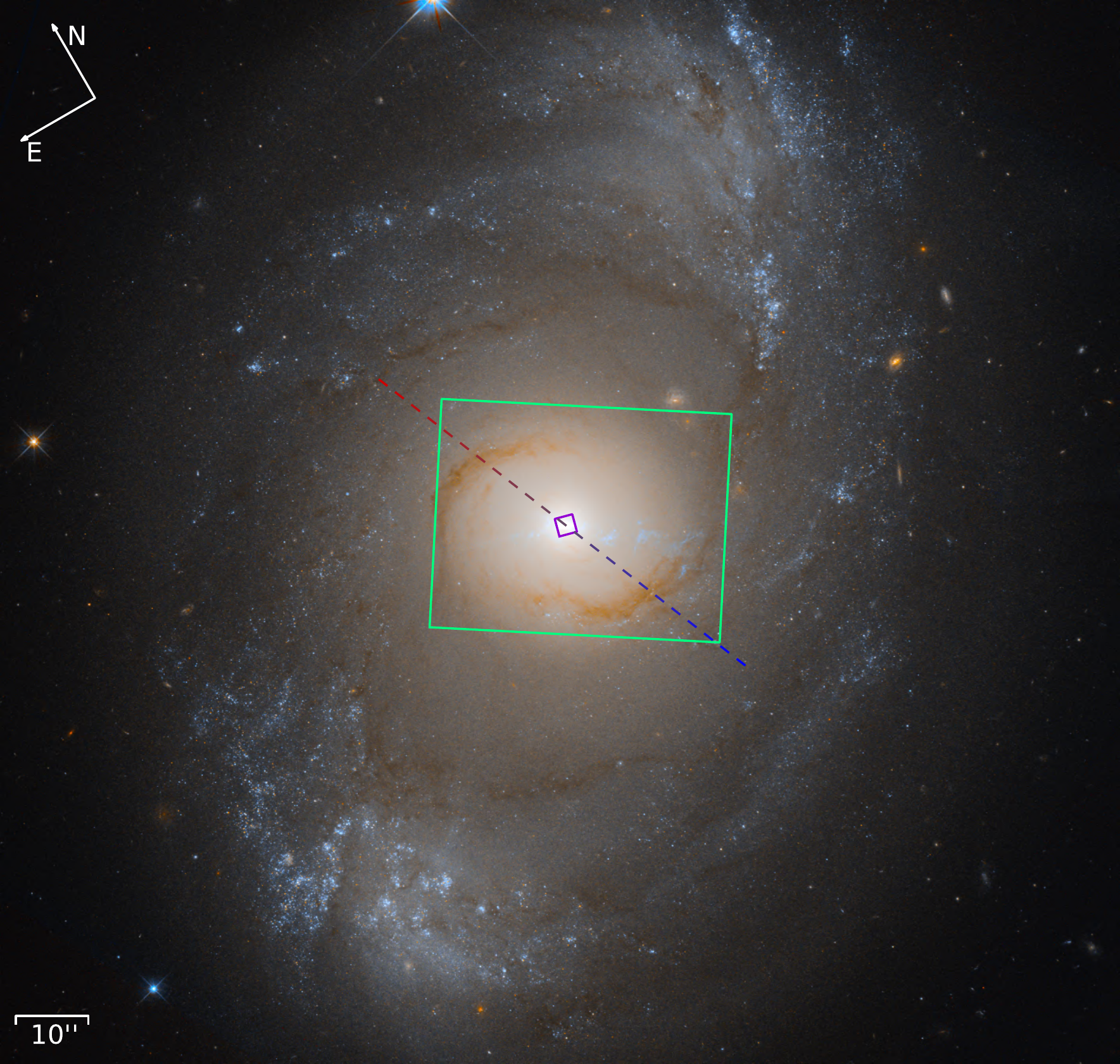}
	\caption{\hst\ Wide-Field Camera 3/UVIS optical image of NGC\,4151 through the F350LP, F814W, and F555W filters. North is 30$^{\circ}$ counter-clockwise from up. Small magenta and large green rectangles show the footprints of the NIFS and SAURON kinematic maps, respectively; dashed red-blue line shows the orientation of the kinematic major axis. Image credit: Judy Schmidt. }
	\label{fig:n4151}
\end{figure}

NGC\,4151 has a \cite{1964rcbg.book.....D} classification\footnote{NASA/IPAC Extragalactic Database; https://ned.ipac.caltech.edu/} of (R?)SAB(rs)ab, having a possible outer ring, a weak bar, an inner ring and inner spiral structure, and tightly wound spiral arms   (Figure~\ref{fig:n4151}).  The large round component that has often been identified as the bulge of the galaxy is actually a barlens, and the true bulge is significantly more compact (cf.\ \citealt{2018ApJ...864..146B}). It is located at $\alpha=182.6357$, $\delta=+39.4058$ in the direction of the constellation Canes Venatici.

\subsection{Integral Field Spectroscopy}

\subsubsection{Gemini NIFS}
NGC\,4151 was observed with the Near-infrared Integral Field Spectrograph \cite[NIFS;][]{2003SPIE.4841.1581M}  on the Gemini North telescope on 2008 February 16-17 and 19-24 (see Table~\ref{tab:observations}). We retrieved the raw data from the Gemini Archive (GN-2008A-Q-41, PI: C.\ Onken). 

NIFS is an image-slicer style integral field unit (IFU) with 29 slices across the $3\farcs0 \times 3\farcs0$ field of view (FOV). The observations of NGC\,4151 were taken in the $H-$band with the H\_G5604 grating coupled with the JH\_G0602 filter, covering the spectral range $\sim 14900-18000$\,\AA\ with R$\approx$5290 and a dispersion of 1.6\,\AA\,pix$^{-1}$. The instrumental resolution was measured to have FWHM = $3.2$\,\AA\,pix$^{-1}$. The data were acquired with the instrument rotated to a position angle (PA) of $-15^{\circ}$ east of north. The Altair AO system  \citep{1998SPIE.3353..488H} was used with the bright AGN serving as a natural guide ``star''. 

\begin{deluxetable}{ccccc}
\renewcommand{\arraystretch}{2}
\tablecolumns{5}
\tablewidth{0pt}
\tablecaption{Gemini NIFS Observations}
\tablehead{
\colhead{Date} &
\colhead{Target} &
\colhead{Type} &
\colhead{Exp.\ Time} &
\colhead{\# Obs.} \\
\colhead{(Feb.\ 2008)} &
\colhead{} &
\colhead{} &
\colhead{(s)} &
\colhead{}
}
\startdata
16	&	NGC 4151	&	Science	&	120	&	54	\\
	&	HD 98152	&	Telluric (A0)	&	40	&	12	\\
	&	HD 116405	&	Telluric (A0)	&	30	&	4	\\
17	&	NGC 4151	&	Science	&	120	&	36	\\
	&	HD 98152	&	Telluric (A0)	&	40	&	8	\\
	&	HD 116405	&	Telluric (A0)	&	30	&	8	\\
19	&	NGC 4151	&	Science	&	120	&	47	\\
	&	HD 98152	&	Telluric (A0)	&	40	&	4	\\
	&	HD 116405	&	Telluric (A0)	&	30	&	12	\\
20	&	NGC 4151	&	Science	&	120	&	19	\\
	&	HD 98152	&	Telluric (A0)	&	40	&	8	\\
	&	HD 116405	&	Telluric (A0)	&	30	&	4	\\
21	&	NGC 4151	&	Science	&	120	&	24	\\
	&	HD 98152	&	Telluric (A0)	&	40	&	8	\\
	&	HD 116405	&	Telluric (A0)	&	30	&	4	\\
22	&	NGC 4151	&	Science	&	120	&	18	\\
	&	HD 98152	&	Telluric (A0)	&	40	&	4	\\
	&	HD 116405	&	Telluric (A0)	&	30	&	4	\\
23	&	NGC 4151	&	Science	&	120	&	36	\\
	&	HD 98152	&	Telluric (A0)	&	40	&	4	\\
	&	HD 116405	&	Telluric (A0)	&	30	&	8	\\
	&	HIP 60145	&	Template (M0)	&	5.3	&	4	\\
24	&	NGC 4151	&	Science	&	120	&	18	\\
	&	HD 98152	&	Telluric (A0)	&	40	&	4	\\
	&	HD 116405	&	Telluric (A0)	&	30	&	4	\\
	&	HD 35833	&	Template (G0)	&	5.3	&	4	\\
	&	HD 40280	&	Template (K0\textrm{III})	&	5.3	&	\textcolor{white}{.}4
\label{tab:observations}
\enddata

\end{deluxetable}

Each exposure of NGC\,4151 had a length of 120\,s, and 252 individual exposures were obtained during the 8 nights of observations. The data quality and weather conditions were good; only 3\% of the data did not meet the expectations in quality checks such as cloud cover, water vapor/transparency, and background counts indicated by the PI prior to the observations. At all times the data were marked as usable by the Gemini staff's quality assessment. The airmass was rarely above 1.5 (only 5\% of the time), and the average airmass was 1.2. 

Observations of Galactic stars were also collected to produce telluric spectra that quantify the absorption of light from molecules in Earth's atmosphere (most notably water vapor). Two A0\textrm{V} stars were used for these telluric observations, HD\,98152 and HD\,116405, with exposure times of 40\,s and 30\,s, respectively. They were observed throughout each of the eight nights, interspersed between the science observations in order to monitor the varying telluric feature strengths due to changing airmass and weather conditions. 

G, K, and M stars dominate the near-IR stellar emission of galaxies, and the high luminosities of giant stars, in particular, are responsible for the bulk of the stellar emission at these wavelengths. Three giant stars --- HD\,35833 (G0), HD\,40280 (K0\textrm{III}), and HIP\,60145 (M0) --- were also observed on Feb.\ 23 and 24 to serve as velocity templates for assistance in the interpretation of the NGC\,4151 spectra. For each of the three stars, four 5.3\,s exposures were obtained in a single night. 

Observations of NGC\,4151 and the telluric and velocity template stars were typically obtained in an object-sky-object dithering pattern to allow for sky subtraction. The telescope was slewed $\sim$200\arcsec\ to the side of the FOV for the NGC\,4151 data ($\sim$5\arcsec\ for telluric and velocity template stars) before re-centering on the target for the next exposure. The AO was turned on for the science data and the telluric stars, but not for the velocity templates. 

Reduction of the data generally followed the NIFS reduction pipeline created by Tracy Beck and Richard McDermid for  IRAF\footnote{IRAF is distributed by the National Optical Astronomy Observatory, which is operated by the Association of Universities for Research in Astronomy (AURA) under a cooperative agreement with the National Science Foundation.}. Many of the basic reduction procedures are the same as those employed in the original study of these data by \cite{2014ApJ...791...37O}, but we highlight the improvements we have made to the reductions below.

In particular, telluric spectra were created from the telluric star observations using the software {\fontfamily{qcr}\selectfont xtellcor} \citep{2003PASP..115..389V}. The telluric star spectra consist of stellar continuum and absorption with telluric features superimposed on top.  {\fontfamily{qcr}\selectfont xtellcor} uses a high-resolution, synthetic spectrum of Vega to model the spectra of A0V stars and isolate the stellar features from the telluric features. Each telluric star spectrum was fed to {\fontfamily{qcr}\selectfont xtellcor} along with the B and V magnitudes of the star (8.98 and 8.93 for HD\,98152, 8.27 and 8.34 for HD\,116405).\footnote{Magnitudes of the telluric stars were retrieved from SIMBAD.}  Our telluric standard stars, while being the same spectral type as Vega, had different absorption line widths in their spectra than those of Vega, and so {\fontfamily{qcr}\selectfont xtellcor} scaled and blurred the model of Vega to better match the intrinsic stellar absorption features. The best-fit model was then subtracted from the observed spectrum of the telluric standard, leaving only the telluric absorption spectrum.  This process differs slightly from that of  \cite{2014ApJ...791...37O}, who instead employed the {\fontfamily{qcr}\selectfont nffixa0} IRAF script written by Peter McGregor, in which the stellar absorption lines of the telluric standards were individually fit with Voigt profiles and removed in order to isolate the features arising from Earth's atmosphere. In both cases, the execution of multiple telluric template observing blocks each night allowed individual frames of NGC\,4151 to be corrected using the telluric template spectrum obtained closest in time. 

After telluric correction, all individual science frames  were then reformatted into three-dimensional data cubes, drizzling and rectifying the 29 spectral slices of each exposure into a $60 \times 62$ spaxel (spatial pixel) data cube with a spatial sampling of $0\farcs05\times0\farcs05$, preserving more of the spatial resolution than \cite{2014ApJ...791...37O} where $0\farcs2\times0\farcs2$ was adopted instead. For the velocity template stars, a one-dimensional spectrum was then extracted from the data cube by summing all the spectra contained within a circular spatial aperture. The four individual spectra for each star were then median combined.

For observations of NGC\,4151, we ensured the wavelength axis was consistent among all the cubes by adopting a fixed dispersion of 1.6\,\AA\,pix$^{-1}$ and identical starting wavelengths for each cube as they were rectified. We then assessed the FWHM of the point spread function (PSF) for each cube and implemented a seeing cut at FWHM = $0\farcs25$, thus rejecting  21 of the 252 individual observations with the poorest spatial resolution (most of which had been observed in the last two nights of observations). This left 231 individual cubes that were aligned and median combined into a final data cube.  

Variance information for each data frame was recorded at the time of observation, but the current version of the NIFS reduction pipeline does not carry this information through to the end of the process. Thus, we developed a method for quantifying the adjustments to the variance at each remaining step in the pipeline and producing a final noise cube to match the final data cube. The inclusion of propagated errors is an improvement on the reduction methods of \cite{2014ApJ...791...37O}, where a constant fractional error of 2\% was assumed. 

\subsubsection{WHT SAURON}
An additional data cube with wider-field IFU observations of NGC\,4151 was provided by Eric Emsellem.  The observations were collected as part of the study by \cite{2007MNRAS.379.1249D}, where integral field spectroscopy from the SAURON IFU was used to examine the kinematics of the stellar and gaseous components of a sample of galaxies including NGC\,4151. Three 30\,min exposures were collected on the 4.2-m William Herschel Telescope (WHT) in La Palma, Spain in 2004 March. SAURON is a lenslet system rather than an image-slicer, and the IFU has a field-of-view of $33\arcsec\times41\arcsec$ with a spatial resolution of 0\farcs94, resulting in $\sim$1500 spectra per pointing. The spectral range is in the optical, covering 4825-5380\,\AA, and the instrument has a spectral resolution of 4.2\,\AA\,pix$^{-1}$. 

The SAURON data were reduced with the {\fontfamily{qcr}\selectfont XSAURON} software and SAURON pipeline \citep{2001MNRAS.326...23B,2002MNRAS.329..513D}. In this process, a bias and dark subtraction was performed, followed by wavelength calibration, flat-fielding, cosmic ray correction, sky subtraction, and flux calibration. The three individual cubes were then aligned and drizzled onto a square grid with a spatial sampling of $0\farcs8\times0\farcs8$. The PSF of the final data cube has FWHM = $2\farcs0\pm0\farcs3$. For further information on the data, we refer the reader to \cite{2007MNRAS.379.1249D}.

\subsection{Photometry}\label{phot}

High-spatial resolution {\it Hubble Space Telescope} (\hst) imaging of NGC\,4151 had been previously acquired with the Wide Field Camera 3 (WFC3) through the F547M (medium-$V$; GO-11661, PI: M.\ Bentz) and F160W ($H$; GO-13765, PI: B.\ Peterson) filters.

Details of the reduction and processing of the F547M images are described in \citet{2018ApJ...864..146B}, but in brief, four images through the F547M filter were acquired in a single orbit on 2010 July 3 with a two-point dither pattern to cover the central gap between the detectors.  At each point in the dither pattern, a short exposure and a long exposure were obtained, allowing for correction of saturation in the nucleus due to the AGN while also providing good sensitivity to the extended host galaxy.  The total exposure time was 2310.0\,s, and the drizzled image covers a field of view of $2\farcm7 \times 2\farcm7$ at a pixel scale of 0\farcs04.

Six individual F160W images were acquired in pairs with equal exposure times and a two-point dither pattern over the course of three orbits on 2015 December 7, 2015 December 28, and 2016 January 5.  The total exposure time was 3317.6\,s, and the drizzled image covers a field of view of $2\farcm3 \times 2\farcm1$ at a pixel scale of 0\farcs1283.

The drizzled F547M and F160W images were each fit separately with \galfit\  \citep{2002AJ....124..266P,2010AJ....139.2097P} to construct two-dimensional surface brightness profiles (see \citealt{2013ApJ...767..149B,2018ApJ...864..146B} for more details).  The surface brightness models allowed the PSF of the central AGN and the background sky to be removed from each image, leaving just the host galaxy for further analysis.

\begin{figure}
\includegraphics[width=3.3in]{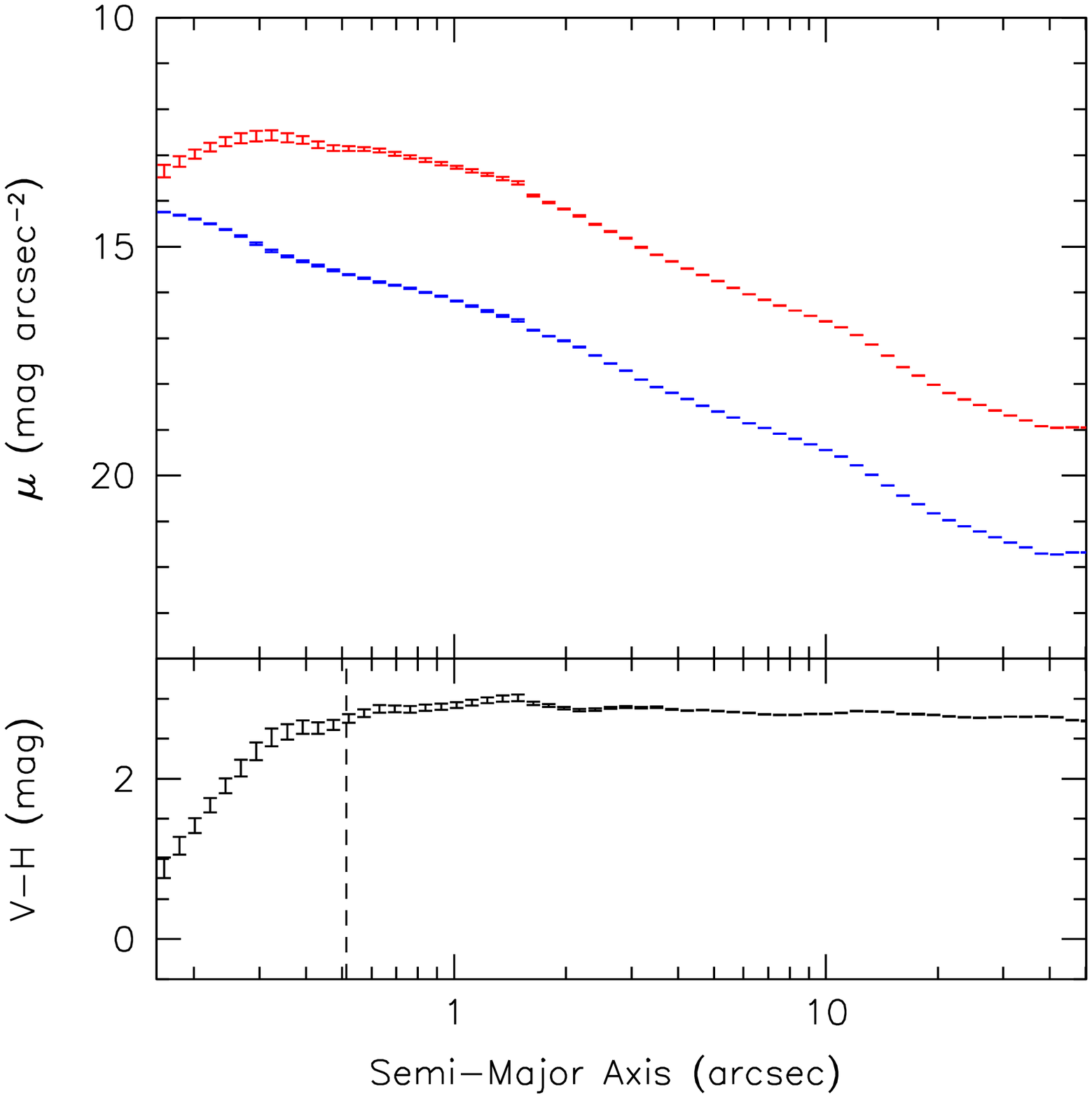}
\caption{$V-$ (blue) and $H-$band (red) surface brightness ({\it top}) and $V-H$ color ({\it bottom}) as a function of radius from the center of NGC\,4151.  The color is relatively constant, except for the innermost 0\farcs5 ($\sim4$ pixels) which are affected by the bright central AGN.}
\label{fig:surfacebrightness}
\end{figure}

To compare the surface brightness profiles of the galaxy in the two filters, we rotated the F160W image to match the orientation of the F547M image, blurred and rebinned the F547M image to match the PSF profile and spatial scale of the F160W image, and then cropped the two images so they were both centered on the AGN and included a common field of view. Comparing the positions of the few resolved sources in common between the two images, we found that this process aligned the two images to an accuracy of $\sim0.5$\,pix at the final spatial scale.  We then used the IRAF task {\sc ellipse} to fit elliptical isophotes to both images and measure their one-dimensional surface brightness profiles in a consistent manner.  Finally, we applied the most recent calibration of the Vegamag zeropoints, corrected for Galactic extinction based on the \citet{2011ApJ...737..103S} recalibration of the \citet{1998ApJ...500..525S} dust map of the Milky Way, and applied small color corrections to account for the differences between a $V-$band filter and F547M, and an  $H-$band filter and F160W.  Figure~\ref{fig:surfacebrightness} ({\it top}) displays the one-dimensional surface brightness profiles in the inner regions of NGC\,4151, with $V$ shown in blue and $H$ shown in red.  The $V-H$ color is displayed in the bottom panel.  The pixels inside a radius of $\sim 0\farcs5$ (equivalent to a radius of about 4 pixels, and denoted by the vertical line in the bottom panel) are affected by residuals from the subtraction of the AGN PSF, but at radii outside this region the $V-H$ color of the galaxy is quite flat across the inner $\sim 1\farcm0$.

\section{Photometric Analysis} \label{sec:phot}

\begin{figure}
\includegraphics{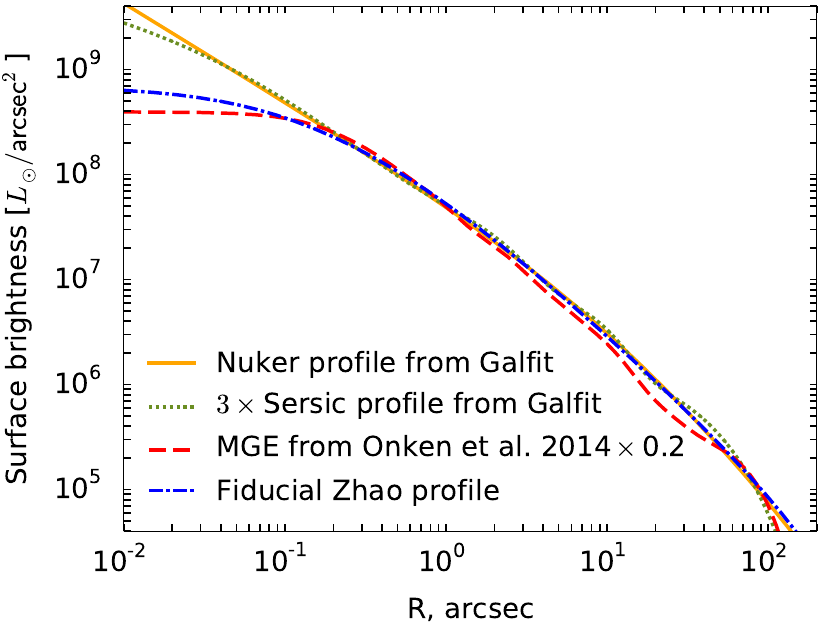}
\caption{Comparison of several parametrizations of the surface brightness profiles. Orange solid line and green dotted lines show the \galfit models to the F547M image with a single Nuker component or 3 S\'ersic components, respectively. Red dashed line is the $H$-band MGE model from \citet{2014ApJ...791...37O} scaled by $0.2$ to match the other profiles, which are based on the $V$-band HST image. Blue dot-dashed line is our fiducial \citet{1996MNRAS.278..488Z} profile, which closely follows the Nuker profile outside 0$\farcs$1 but is less cuspy.
} \label{fig:density_profiles}
\end{figure}

\subsection{Image Decompositions}
The surface brightness of NGC\,4151 in both the WFC3 F547M and F160W images were examined for their appropriateness in constraining the stellar luminosity density within the dynamical models.  The F547M image shows clear regions of dust extinction and excess stellar flux, while the F160W image shows a smoother galaxy profile but at the expense of a lower intrinsic spatial resolution than the F547M image.  

We first tried a Multi-Gaussian Expansion (MGE) analysis
\citep{1994A&A...285..723E,2002MNRAS.333..400C} using the AGN- and sky-subtracted images of the galaxy discussed above. MGE surface-brightness decomposition begins by finding the center then binning the field in angle and radius. Photometry is measured for each point on the grid, and then nested two-dimensional Gaussian profiles with variable axis ratios and central intensities are fit to this two-dimensional intensity map. During this process, we required that the PAs of any elongation in the Gaussians aligned with the kinematic PA determined previously  (this restriction is mandated by our choice of axisymmetric modeling strategy). Although the observed velocity field shows twisting and misalignment with the photometric PA, indicative of the presence of a bar, we defer the study of bar models to the future.

The parameters of the best-fit nested Gaussian profiles are described by the intensity in counts, the width in pixels ($\sigma$), and axis ratio ($q$).  These parameters can then be converted into surface density in $L_{\odot}$\,pc$^{-2}$ and $\sigma$ in arcsec. For both the F547M and the F160W images, the MGE solutions provided a surface brightness profile that included a lot of bumps and other structures, rather than a smooth increase in surface brightness from larger radii into the galaxy center.

We then examined the potential of using the surface brightness components that had been fit to the images with \galfit\ in the process of subtracting the AGN and sky.  These surface brightness components included an exponential disk, S\'ersic bulge, and an elongated S\'ersic component to represent the weak bar, with the central positions of all the galaxy components tied together, and the position angles of the central galaxy components (not including the disk) set to the kinematic position angle of the inner galaxy.  While smoother than the MGE profile, a significant amount of structure was still visible in the surface brightness profile for both the F547M and the F160W image.

Finally, we also examined the results of fitting a Nuker profile to the images using \galfit, to promote smoothness in the surface brightness profile while still retaining some flexibility to account for the complicated morphology.  Figure~\ref{fig:density_profiles} demonstrates that the Nuker fit to the F547M image is very close to the combination of several S\'ersic profiles, but is smoother. However, both these profiles overpredict the amount of stellar light in the central $0\farcs1$ -- the region so dominated by the AGN emission that any residual stellar light is poorly constrained. A comparison with the MGE model from \citet{2014ApJ...791...37O}, suitably scaled in normalization, shows that the latter has a central core with the radius of the innermost Gaussian component $\sim 0\farcs11$, while the Nuker model has a cusp. As will be shown later in Section~\ref{sec:models}, a cuspy profile results in a best-fit black hole mass of zero, which is clearly unphysical for this bright AGN. We therefore constructed another model described by the \citet{1996MNRAS.278..488Z} profile, which closely follows the Nuker model except the innermost $0\farcs1$ -- the region dominated by the AGN emission.  We note that both Nuker and Zhao profiles are equally flexible and can represent cuspy or cored systems for different choices of parameters; here we used the former parametrization for the cuspy profile and the latter for the cored one. These two extreme cases likely encompass the true range of possible density profiles for this galaxy. The 3d luminosity density for our adopted Zhao profile is \begin{equation}
j(r)=j_0\,(r/r_0)^{-\gamma}\,\big[1+(r/r_0)^\alpha\big]^{(\gamma-\beta)/\alpha},
\end{equation}
with $\gamma=0$, $\beta=2.47$, $\alpha=0.65$, $r_0=0\farcs113$, and $j_0=4.08\times10^6\,L_\odot\,\mathrm{kpc}^{-3}$.
We note that although $\gamma=0$ formally corresponds to a cored profile, the zero logarithmic slope is achieved only asymptotically, and in practice the density remains weakly cuspy at the resolution limit, unlike the MGE parametrization, which is much more obviously cored within $0\farcs1$.

\subsection{Mass-to-Light Ratio}   \label{sec:mass_to_light}

Determination of how mass traces light is integral to understanding the gravitational potential of the galaxy for modeling.  Based on OSUBSGS $H-$band imaging combined with SDSS $g-$ and $i-$band imaging, \cite{2014ApJ...791...37O} produced color-color maps of NGC\,4151 and found very flat colors across the galaxy bulge with average values of $g - i=1.1 \pm 0.1$\,mag and $i - H=2.4 \pm 0.2$\,mag. 
This is in line with our findings based on \hst\ F547M and F160W imaging as described in Section~\ref{phot}, where we also find a very flat color across the galaxy bulge, with $V-H=2.9\pm0.1$\,mag.

\cite{2014ApJ...791...37O} adopted the models of  \cite{2009MNRAS.400.1181Z} to predict $\Upsilon_{H} \simeq 0.4 \pm 0.2$\, M$_{\odot}/$L$_{\odot}$.  Comparison of many different prescriptions for estimating $\Upsilon$ by \citet{roediger15}, however, finds that \cite{2009MNRAS.400.1181Z} consistently predicts the lowest values at these galaxy colors.  Furthermore, when compared to the known $\Upsilon$ for resolved stars in M31 \citep{telford20}, the prescriptions of  \cite{2009MNRAS.400.1181Z} seem to be biased low at these colors.  

Based on the analysis by \citet{roediger15}, we therefore adopt $\Upsilon_{H} \simeq 0.7 \pm 0.1$\,M$_{\odot}/$L$_{\odot}$ and $\Upsilon_{V} \simeq 2.9 \pm 0.3$\,M$_{\odot}/$L$_{\odot}$ as our best  estimates for the $H-$ and $V-$band stellar mass-to-light ratios in the nucleus of NGC\,4151.

\subsection{Galaxy Inclination Angle} \label{sec:inclination}

Observations of HI 21\,cm emission from the large-scale disk of NGC\,4151 led \citet{davies1973} to report an inclination of $26^{\circ}\pm8^{\circ}$.  \cite{1975ApJ...200..567S} examined the ratio of the minor to major axis using density tracings of wide-field photographic plate images and found the outermost isophotes of the galaxy to suggest an inclination of $21^{\circ} \pm 5^{\circ}$.  The nearly circular shape of the galaxy disk in the HI 21\,cm maps presented by \citet{1977A&A....57...97B} also support a face-on orientation.

This is at odds with what has generally been determined from optical imaging alone.  For example, surface brightness decomposition of optical ground-based imaging with a modest FOV generally provides a minor-to-major axis ratio of $\sim 0.7$ for the galaxy disk (e.g., \citealt{2009ApJ...697..160B}), suggesting an inclination of 46$^{\circ}$. However, as argued by \citet{davies1973}, this elongated structure is likely to be more representative of the weak galaxy bar than the disk, although they also note that there are difficulties with this interpretation as well. 

Thus the inclination of NGC\,4151 is somewhat uncertain (\citealt{1975ApJ...200..567S} go so far as to describe NGC\,4151 as ``pathological''!), but is likely to be in the range of $20-45^{\circ}$.

For a nearly face-on orientation, it is very difficult to estimate the intrinsic flattening $q$ from the photometry (even a very flat disk would still appear nearly round in projection). Therefore, our strategy is to explore the parameter space of the inclination angle $i$ and intrinsic flattening $q$ independently. Specifically, we first take the 1d surface brightness profile $I(R)$ and deproject it under the assumption of spherical symmetry, obtaining the luminosity density profile 
\begin{equation*}
j(r) = -\frac{1}{\pi} \int_r^\infty \frac{dR}{\sqrt{R^2-r^2}}\,\frac{dI}{dR}.
\end{equation*}
We then assume that the intrinsic luminosity density is actually axisymmetric and ellipsoidally stratified, with the axis ratio $q\equiv z/R$ being a free parameter in the model, where $q=1$ is spherical. If such a density profile is projected face-on, the result would be identical to $I(R)$, while for a nonzero inclination angle $i$, the projected minor axis is $q'=\sqrt{1-(1-q^2)\,\sin^2 i}$\; times smaller than the projected major axis. To preserve the angularly-averaged surface brightness profile, we multiply the scale radius of the 3d profile by $s \equiv 1/\sqrt{q'}$, so that the geometric mean of the projected major and minor axes is the same as in the spherical model. For instance, with an inclination $i=30^\circ$ and flattening $q=0.25$, the projected major and minor axes are $1.07$ and $0.935$ times the original scale radius of the one-dimensional surface brightness profile. We believe that the extra flexibility allowed by independent variation of flattening and inclination (with additional constraints coming from kinematics) is more important than an accurate reproduction of the density profile \textit{per se}.

\subsection{Galaxy Distance} \label{moddistance}

When performing SD modeling, the distance to the galaxy is a large source of uncertainty in determining \mbh. The scale or radii of the galactic features being probed and the value of \mbh scale linearly with the assumed distance.  Based on the information available at the time, \cite{2014ApJ...791...37O} adopted a recessional velocity distance measurement of 13.9\,Mpc from \cite{1992MNRAS.259..369P}, cautioning that this distance was highly uncertain but that no better estimates were currently available.

Indeed, the only other information available was a Tully-Fisher distance to NGC\,4151 that was reported by \citet{2009AJ....138..323T} as 3.9\,Mpc and is highly-discrepant from the group-averaged distance to NGC\,4151 of $11.2\pm1.1$\,Mpc.  \citet{2013ApJ...767..149B} recalculated the group-averaged distance to NGC\,4151, excluding NGC\,4151 itself, and reported a distance of $16.6\pm3.3$\,Mpc, albeit based on only three galaxies.

Since then, additional studies have worked to pin down the distance more accurately.  \cite{2014Natur.515..528H} measured the physical size of the inner radius of the  dusty torus using broad-band RM and the angular size of the torus with near-infrared interferometry, reporting a dust-parallax distance of $19.0^{+2.4}_{-2.6}$\,Mpc to NGC\,4151.  \citet{tsvetkov19} reported the discovery of a Type IIP supernova in NGC\,4151.  Adopting a standardizable candle approach, they find a distance of $16.6\pm1.1$\,Mpc, but they also report a distance of $20.0\pm1.6$\,Mpc if they instead adopt an expanding photosphere analysis.

Most recently, a Cepheid distance to NGC\,4151 has been obtained with \hst\ WFC3 optical and near-infrared imaging (GO-13765, PI: B.\ Peterson), resulting in a distance of 15.8$\pm$0.4\, Mpc \citep{yuan2020}, which we have  adopted for our modeling.  At this distance, 1\arcsec\ corresponds to 77 pc.

\subsection{Point Spread Function} \label{modpsf}

To characterize the PSF of our NIFS datacube, we took a slice at a wavelength corresponding to an AGN broad emission line and subtracted off a slice at a wavelength corresponding to the stellar and AGN continuum, in effect creating narrowband images with an on-band image and an off-band image. The residual image thus contains an image of the unresolved AGN point source at a wavelength corresponding to a broad emission line, allowing for characterization of the PSF.  

Using \galfit, we modeled the PSF image with multiple two-dimensional Gaussian components, each having common centers but different widths ($\sigma$) and flux contributions. The final three-component model with circular Gaussian components is shown in Table~\ref{tab:psf}. This three-component model accurately matches the wings and the core of the PSF. If only two components were used, the fit was not as good but the values agreed more closely with what was found by \cite{2014ApJ...791...37O}. 

For the SAURON datacube, the PSF was characterized by \cite{2007MNRAS.379.1249D} as a single Gaussian component with width $\sigma=2\farcs0$. 

\begin{deluxetable}{ccc}
\renewcommand{\arraystretch}{2}
\tablecolumns{3}
\tablewidth{0pt}
\tablecaption{NIFS PSF Characterization}
\tablehead{
\colhead{Component} &
\colhead{Weight} &
\colhead{$\sigma$ (\arcsec)}
}
\startdata
1 & 0.339 & 0.045\\
2 & 0.306 & 0.082\\ 
3 & 0.356 & 0.409
\label{tab:psf}
\enddata

\end{deluxetable}
\bigskip

\section{Kinematic Analysis} \label{sec:kin}

In the optical and near-infrared, the spectra of galaxies are dominated by the summation of spectra of luminous stars. The line-of-sight velocity distribution (LOSVD) gives the full range of projected stellar kinematics along a particular line-of-sight. To describe the bulk motions of unresolved populations of stars at each spatial position, the convolution kernel necessary to transform stellar template spectra into the absorption profiles of observed galaxy spectra must be determined. In addition to a central wavelength shift indicating typical velocity $V$, the detailed shapes of the convolution kernels can be approximated with higher-order Gauss-Hermite terms, including the width of the profile $\sigma$ (related to the velocity dispersion), $h_3$ (related to skewness), $h_4$ (related to kurtosis), $h_5$, and $h_6$. Data with S/N$>$30 is generally necessary for making sure that the higher order moments of the Gauss-Hermite polynomials can be constrained \citep{1994MNRAS.269..785B}.  We employed the Penalized Pixel-Fitting method (pPXF; \citealt{2004PASP..116..138C, 2017MNRAS.466..798C}), which was developed to constrain the Gauss-Hermite approximation to absorption line profiles through determination of the shifting and blurring required to match a stellar absorption template to an observed galaxy spectrum.

\subsection{Kinematic Position Angle}

The kinematic position angle was derived using the method of \cite{2006MNRAS.366..787K}.  The kinematic measurements derived from an initial pPXF run were smoothed and bi-symmetrized before determining the best-fit systemic velocity along with the kinematic position angle of the system as measured from the y-axis of the data cube. This angle represents the line of maximum galaxy rotation, perpendicular to the axis of rotation, and for our kinematics is 37.2$^{\circ}$ counterclockwise from the y-axis. 

The photometric major axis of the galaxy is somewhat uncertain. At intermediate radii ($\gtrsim 30\arcsec$), the galaxy is visibly elongated in a direction almost perpendicular to the kinematic major axis (see Figure~\ref{fig:n4151}), while at smaller radii (within the footprint of the SAURON dataset, shown by a green rectangle), its elongation is still offset by some $30^\circ$ from the kinematic major axis. Furthermore, 21\,cm imaging of the galaxy shows a nearly circular gas disk extending several arcmin beyond the high surface brightness stellar features, with a position angle of $26^{\circ}$ \citep{davies1973}. The intrinsic galaxy shape is thus non-axisymmetric and radius-dependent. However, in the present study we do not attempt to model this complex morphology and instead assume an intrinsically axisymmetric shape. The use of axisymmetric models requires that the kinematic and photometric major axes are aligned, which motivates the constraint on the position angle during the photometric fit.

\subsection{Voronoi Binning}

Next, we employed the adaptive binning method of \cite{2003MNRAS.342..345C}. This binning procedure is based on \citet{1908C...134....198V} tessellations and assigns adjacent spaxels into bins in a scheme that approximates constant signal-to-noise (S/N) across the FOV. Bins that are similar in shape to the spaxels are prioritized, i.e.\ not overly irregularly shaped, even at the expense of some irregularity in the S/N. This Voronoi binning method preserves the high spatial resolution at the center of the galaxy where the S/N is highest while adaptively binning the spaxels at the edges of the field to increase their collective S/N. 

The data and noise cubes were first cropped, ensuring that the center of the FOV was defined by the centroid of the AGN position, and that spaxels at the edges of the cube with very low counts due to  dithering between exposures were rejected. This left us with 49$\times$49 spaxels, with the odd number of spaxels in each dimension ensuring that the central spaxel contained the AGN.  We then produced both a slice of the data cube (the signal) and a slice of the noise cube that well-represented the cubes as a whole, focusing on the region between the 16200\,\AA\, CO (6-3) bandhead and the strong [FeII] emission line at 16440\,\AA.  These signal and noise slices were then used to assign the individual spaxels to bins.

We identified the Voronoi bin pattern for only one-quarter of the FOV, bounded by the kinematic major and minor axes, and then we replicated the bin pattern for the other three quadrants of the FOV.  This scheme ensures that each bin has symmetric analogs in the other quadrants of the FOV, while the central spaxel containing the AGN was assigned to its own single bin. With this approach, we divided the FOV into 165 total bins.  

Once the Voronoi bin assignments were determined, the spaxels that were assigned to each particular bin were combined for both the data cube and the noise cube. For all the spaxels assigned to a particular bin, the spectra from the data cube were co-added, channel by channel, and the spectra from the noise cube were added in quadrature.  Voronoi binning the NGC\,4151 integral field spectroscopy is an improvement of the analysis over that of \cite{2014ApJ...791...37O}, where the cubes were rectified with a constant spatial sampling of 0\farcs2$\times$0\farcs2  across the FOV, thus degrading the spatial resolution in the crucial region near the black hole sphere of influence.

\subsection{NIFS Kinematic Fits}  \label{sec:nifs_kinem}

The NIFS spectra cover the wavelengths $\sim 15100-17700$\,\AA. Notable features within the wavelength range include forbidden Fe emission on the blue side, including the strongest line at 16440\,\AA, and hydrogen Brackett 11 and 10 lines on the red side at 16811\,\AA\ and 17367\,\AA, respectively. We focused on the wavelength range $\sim 15200-16400$\,\AA, which included the $\sim$16200\,\AA\, CO (6-3) bandhead and blueward absorption.  The spectra were fit within the wavelength ranges $\sim 15150-15330$, $15450-15500$, $15530-16000$, $16060-16130$, and $16180-16350$\,\AA, the same as \cite{2014ApJ...791...37O}. These ranges satisfactorily masked the strong emission lines while focusing the analysis on the strongest expected stellar absorption features.

As LOSVDs can be described by Gauss-Hermite polynomials, when the noise of the spectra is high or the data are undersampled with a low velocity resolution, the proper use of pPXF is to bias or penalize higher order Gauss-Hermite terms toward zero, defaulting them to more Gaussian shapes. In the original study, \citet{2014ApJ...791...37O} adopted no penalty.  Based on the details of the NIFS integral field spectroscopy, however, we adopted a {\fontfamily{qcr}\selectfont BIAS} of 0.3.  We also examined the largest changes in kinematic measurements between {\fontfamily{qcr}\selectfont BIAS} = 0.3 and  {\fontfamily{qcr}\selectfont BIAS} = 0 to determine which bins produce unreliable kinematic measurements that might need to be masked during the modeling process. 

While convolving the velocity template spectra to match the observations, pPXF provides the option to include Legendre polynomials (\citealt{1920WW}) to improve the fit to the continuum. The additive polynomials ({\fontfamily{qcr}\selectfont DEGREE}) can mediate some of the effects of template mismatch (where poor fits of kinematics are obtained due to templates that incompletely represent the observations) and imperfect sky subtraction in the data reduction, while the multiplicative polynomials ({\fontfamily{qcr}\selectfont MDEGREE}) can aid with slight imperfections in the flux calibration and reduce or remove the need for reddening correction.  We found a good balance between the weights of the various polynomial components and the goodness of fits to the galaxy spectra when adopting 2nd order additive and 2nd order multiplicative Legendre polynomials, identical to what was found by \cite{2014ApJ...791...37O}. These low-order polynomials improved the overall shape of the best fits to the spectra without introducing localized fluctuations that might have inhibited our ability to constrain the stellar kinematics.

We adopted the point-symmetric two-sided pPXF fitting of our spectra, in which two identically-shaped bins symmetric across the center ($x,y$ and $-x,-y$) are fed to pPXF and fitted simultaneously. Point-symmetric fitting improves the S/N of the data and provides symmetric measurements that utilize the full information contained in each of the two spectra.  This is another improvement in the analysis over that of \cite{2014ApJ...791...37O}, where the bins were fit independently and then the output measurements ($V, \sigma, h_3, h_4$)  were four-fold symmetrized (also known as bisymmetrization, described below) instead and their errors from individual bins were added in quadrature and divided by two. We opted not to perform such bisymmetrization, in which spectra from four bins at positions ($\pm x,\pm y$) are fitted simultaneously, for two reasons. First, bisymmetrized kinematic maps hide any non-axisymmetric features (although we currently explored only axisymmetric models, in the future we plan to extend the analysis to a more general geometry). Second, in the point-symmetrized case we have essentially two independent variants of kinematic maps (one in the opposing pair of quadrants $x>0,y>0$ and $x<0,y<0$, the other -- in the second pair $x>0,y<0$ and $x<0,y>0$; here $x$ and $y$ are coordinates aligned with the kinematic major axis). Even though for most models we fitted both halves of the dataset simultaneously (and hence the model maps, which are bisymmetric by construction, lie halfway between the two variants of the observed maps), we can fit the two variants separately to quantify the systematic differences in the model parameters arising from the asymmetries in the data -- this is explored later in Section~\ref{sec:model_grid} and in Figure~\ref{fig:chi2_nifs_half}.

In our tests, we found that the G star was only used in 20\% of the best-fit spectra, and even when it was used its contribution had a low weight, and so we omitted it as a template. The M star contributed 98\% of the time and the K star contributed 97\% of the time, and so we included both the K and M templates in our final fit. This is a slight difference from the method of \cite{2014ApJ...791...37O}, who used only the M star template in the final kinematic fits. 

\begin{figure*}
	\includegraphics{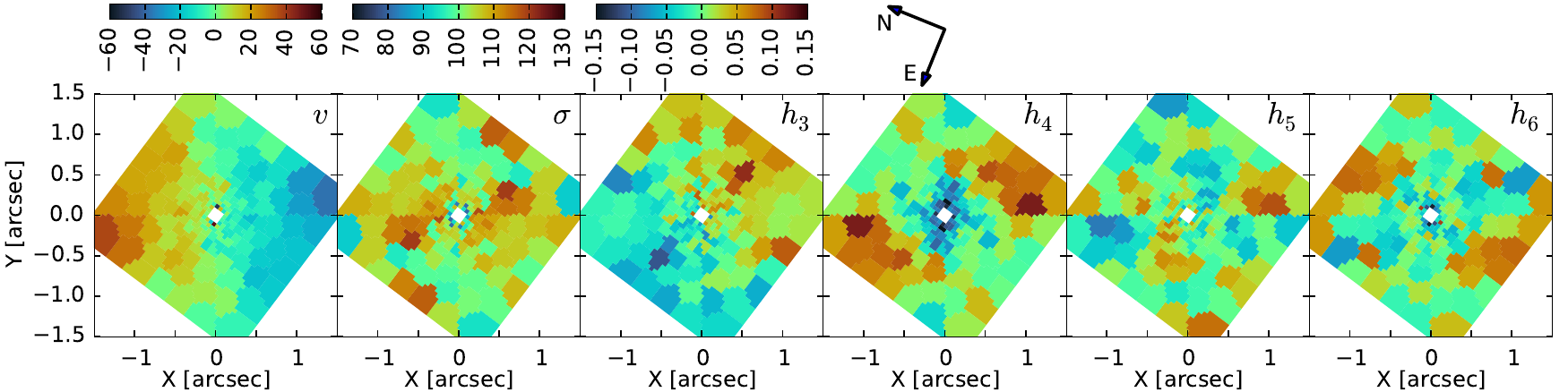}
	\caption{Kinematics measured from the NIFS data. $X$ axis is aligned with the kinematic major axis (position angle $202^\circ$ from North through East). The color bar for Gauss-Hermite coefficients $h_4-h_6$ is the same as for $h_3$ }
	\label{fig:nifs}
\end{figure*}

With all of these adopted parameters, we determined the final stellar kinematics for the central region of NGC\,4151 as shown in Figure~\ref{fig:nifs}.  We were not able to reliably recover the stellar kinematic signature at the center because of the very bright AGN, so for the NIFS data we omit these bins by masking the central $3\times3$ spaxels. This left 156 bins in our final kinematic maps, which are shown in Figure~\ref{fig:nifs}

In summary, the stellar kinematic maps presented here are based on the following improvements over the analysis procedures of \cite{2014ApJ...791...37O}: (1) an improved telluric absorption correction; (2) the inclusion of noise information carried through the full reduction pipeline (rather than assuming a constant noise of 2\%); (3) the adoption of Voronoi binning to preserve the native spatial resolution near the galaxy nucleus and maintain a constant S/N across the field; (4) the adoption of a penalty term in the implementation of pPXF; and (5) fitting the co-added symmetrized spectra in Voronoi bins with pPXF instead of symmetrizing the pPXF measurements in 0.2\arcsec bins. 

\subsection{SAURON Kinematic Fits}  \label{sec:sauron_kinem}

We also reanalyzed the SAURON data of NGC\,4151. The data cube was provided to us having already been Voronoi binned, with 482 total bins that were not symmetric across the FOV. We followed many of the same procedures as the original analysis carried out by \cite{2007MNRAS.379.1249D}, but we updated a few details, including the use of the MILES stellar template library of \citet{2006MNRAS.371..703S} with the updates and corrections of \citet{2011A&A...532A..95F}. The library contains 985 empirical stellar spectra covering $3525-7500$\,\AA\ with a well-constrained spectral resolution of 2.51\,\AA\ \citep{2011A&A...531A.109B}.  We focused on a subset of 148 stars to consider as templates, selecting those stars that were in common with both the Indo-US and Elodie libraries \citep{2004ApJS..152..251V,2001A&A...369.1048P,1998A&AS..133..221S,1998A&A...338..151K} and spanned spectral types F0 to K7.

\begin{figure}
	\includegraphics{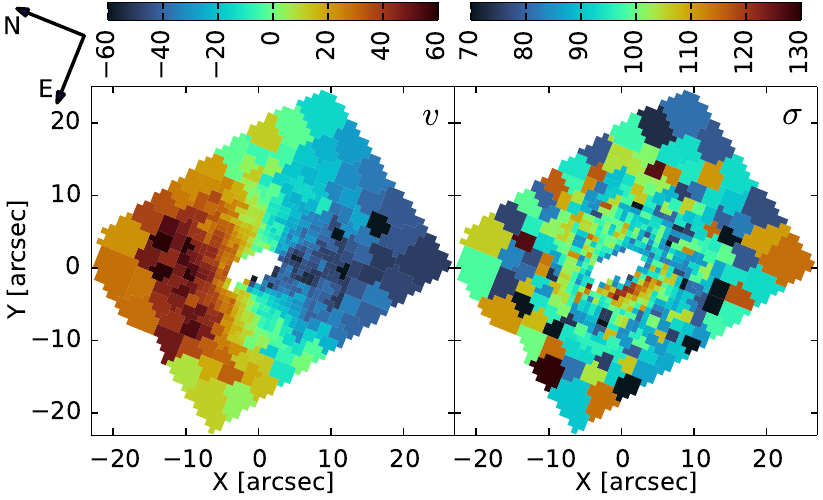}
	\caption{Kinematics measured from the SAURON data. The orientation and the color scales are the same as in Figure~\ref{fig:nifs}. The central region is excluded due to AGN contamination.}
	\label{fig:sauron}
\end{figure}

During the fitting with pPXF, we included the entire spectral range of $4825-5380$\,\AA\  except for small regions around the strong AGN emission lines, including H$\beta$ and [OIII] $\lambda \lambda 4959,5007$\,\AA. Following \citet{2011MNRAS.413..813C}, we determined a single best-fit template spectrum that was then adopted for all the bins, to avoid issues that might arise in the derived kinematics from imperfect velocity calibration of the various template stars in the library. We adopted a {\fontfamily{qcr}\selectfont BIAS} parameter of 0.4 and additive and multiplicative Legendre polynomials of degree 4 and 0, respectively. Because the data cube was already binned with a non-symmetric binning pattern, we were unable to carry out the kinematic fits using the two-sided point-symmetric option, so each bin was fit individually.  Only the first two Gauss-Hermite moments, $V$ and $\sigma$, were included in the fit, with the other terms set to zero. 
We then masked several dozen AGN-contaminated spaxels within a few arcseconds from the center, retaining 447 bins in the final dataset, which is displayed in Figure~\ref{fig:sauron}.

\section{Dynamical Modeling}  \label{sec:models}

\subsection{Method}

The \citet{1979ApJ...232..236S} orbit-superposition method is a standard approach for dynamical modelling of galaxies based on integrated-light stellar kinematics. In the last two decades, several independent implementations of this method have been widely used to determine black hole masses:
the \textsc{Nuker} code \citep{2000AJ....119.1157G,2003ApJ...583...92G,2004MNRAS.353..391T,2009ApJ...693..946S}, 
the \textsc{Leiden} code \citep{1998ApJ...493..613V,1999ApJS..124..383C,2002ApJ...578..787C}, 
the \textsc{MasMod} code \citep{2004ApJ...602...66V,2005ApJ...628..137V}, and 
the \textsc{Heidelberg} code \citep{2008MNRAS.385..647V,2010MNRAS.401.1770V}. 

In this paper, we use yet another recently developed code \textsc{Forstand}\footnote{The code is publicly available, for details see \citet{forstand}.} \citep{2020ApJ...889...39V}. Although it can be applied to galaxies of any geometry, in the present case we limit ourselves to the axisymmetric case. We refer to that paper for a detailed description of the method, and here only briefly summarize the general workflow of orbit-superposition modelling and the particular aspects used in this study of NGC 4151.
\begin{enumerate}
\item First, we determine the 3d luminosity density profile from the surface brightness profile. As explained in Section~\ref{sec:inclination}, we vary the assumed flattening $q$ and inclination $i$ independently, and each combination of these two parameters produces a different 3d density profile (although the spherically averaged profiles are all very similar). The \textit{mass} density profile $\rho(\boldsymbol r)$ is $\Upsilon$ times the \textit{luminosity} density $j(\boldsymbol r)$, where the mass-to-light ratio $\Upsilon$ is another free parameter that is varied at a later stage in the modelling pipeline. We denote the mass density profile with a fixed fiducial value of $\Upsilon_0$ as the ``baseline'' profile $\rho_0$.
\item The total gravitational potential $\Phi(\boldsymbol r)$ contains the contributions from the stars $\Phi_\star$ (which is related to $\rho_0$ by the Poisson equation), the black hole $\Phi_\mathrm{BH} \equiv -GM_\mathrm{BH}/r$, and optionally the dark matter halo $\Phi_\mathrm{h}$, for which we adopt a spherical NFW profile with a scale radius $r_\mathrm{h}$ and peak circular velocity $v_\mathrm{h}$. For simplicity, we fix $r_\mathrm{h}$ at $50\arcsec$ and vary only $v_\mathrm{h}$; since the inner part of the model is self-similar, it is sufficient to vary only the overall density normalization ($\rho\propto v_\mathrm{h}^2$).
\item For each choice of model parameters $i$, $q$, $M_\bullet$, and $v_\mathrm{h}$, we construct an orbit library by integrating $N_\mathrm{orb}$ orbits for 100 dynamical times in the given potential and recording the LOSVDs of each orbit in each Voronoi bin (convolved with the PSF). The initial conditions for the orbits are sampled randomly: the positions are sampled from the stellar density profile, and the velocities -- from a 3d Gaussian distribution with mean and dispersions given by the solution of an axisymmetric Jeans equation for the stellar density in the total potential. The random sampling ensures that all possible orbits supported by the given potential can enter the library, but the results (kinematic fit quality expressed in terms of $\chi^2$) do depend on the specific random realization, because some realizations might contain orbits that are more suitable for reproducing certain kinematic fluctuations caused by observational errors than other realizations. We stress that the scatter in $\chi^2$ values between different orbit libraries with the same model parameters is a general consequence of the highly flexible nature of the method, and the ability to consider different sets of initial conditions does not cause this effect but merely uncovers it. Nevertheless, this stochasticity is clearly undesirable when comparing $\chi^2$ values between models with different parameters, and we use two approaches to reduce its impact: (1) for each series of models with given $i$, $q$, and $v_\mathrm{h}$ but different $M_\bullet$, we use the same set of orbital initial conditions, and (2) we run series of models for several random realizations of the orbital initial conditions and consider the ``ensemble average'' $\chi^2$ contours to determine the range of best-fit model parameters.
\item We then seek a solution for orbit weights that (1) reproduces the 3d density discretized on a $20\times15$ cylindrical grid in the $R,z$ plane, and (2) minimizes the objective function $\mathcal F \equiv \mathcal F_\mathrm{kin} + \mathcal F_\mathrm{reg}$. The first term is the measure of fit quality for the kinematic constraints ($v$, $\sigma$, $h_3\dots h_6$). The second term is the regularization score, which penalizes large variations between orbit weights $\boldsymbol w$; in our case, $\mathcal F_\mathrm{reg} =  \lambda\,N_\mathrm{orb}^{-1}\,\sum_{i=1}^{N_\mathrm{orb}} (w_i/ \overline w)^2$, where the mean orbit weight is $\overline w \equiv M_\star / N_\mathrm{orb}$, and the regularization coefficient $\lambda$ controls the tradeoff between smoothness (for large $\lambda$) and closeness of reproduction of kinematic constraints (for negligible $\lambda$). With a too small value of $\lambda$, there is a risk of overfitting (the model trying to reproduce noise in the data), which is undesirable because $\mathcal F_\mathrm{kin}$ of the best-fit solution exhibits large fluctuations between adjacent points in the parameter space. While a detailed analysis of model sensitivity and fidelity as a function of regularization coefficient is beyond the scope of this study, after some experiments we settle on a value $\lambda=100$ that produces reasonably smooth likelihood surfaces.
\end{enumerate}
The solution for orbit weights in step 4, and the corresponding $\chi^2$ score, are obtained separately for each value of $\Upsilon$, but reusing the same orbit library constructed in step 3, only rescaling the velocity in the model by $\sqrt{\Upsilon/\Upsilon_0}$ (which corresponds to a rescaling of all masses by $\Upsilon/\Upsilon_0$). We start the search from a plausible initial guess for $\Upsilon$ and increase or decrease it by a factor $2^{1/16}\approx 1.04$, until reaching a difference between $\chi^2(\Upsilon)$ and $\chi^2_\mathrm{min}$ greater than 100. The whole process is then repeated from steps 1--3 for a different choice of other model parameters ($i$, $q$, $M_\bullet$, and $v_\mathrm{h}$). In total, we considered a few thousand models, which took $\sim10^3$ CPU hours (the code is parallelized for multi-core architectures, so the wall-clock time is far shorter).

A technical detail worth mentioning is that when using Gauss--Hermite moments as observational constraints, one needs to decompose the LOSVDs of orbits in the model in the same basis, using the observed values $v$ and $\sigma$ as the parameters of the Gaussian--Hermite series for each bin, but expressing the measurement uncertainties $\delta v$, $\delta \sigma$ as uncertainties on the first two coefficients $\delta h_1$, $\delta h_2$, whose measured values are zero by construction. This makes the kinematic objective function $\mathcal F_\mathrm{kin} = \sum_{b=1}^{N_\mathrm{bins}} \sum_{c=1}^6 \big[ (h_{b,c}^\mathrm{model} - h_{b,c}^\mathrm{data}) / \delta h_{b,c} \big]^2$ quadratic in orbit weights and amenable to efficient quadratic optimization solvers. After obtaining the best-fit solution for orbit weights, we then compute the final $\chi^2$ with respect to the original measured values $v, \sigma, h_3\dots h_6$ and their associated uncertainties, which is somewhat different from $\mathcal F_\mathrm{kin}$. Nevertheless, the shapes and locations of the minima are similar for both $\chi^2$ and $\mathcal F_\mathrm{kin}$ as functions of model parameters.

\subsection{Analysis of the Model Grid}  \label{sec:model_grid}

\begin{figure*}
\includegraphics{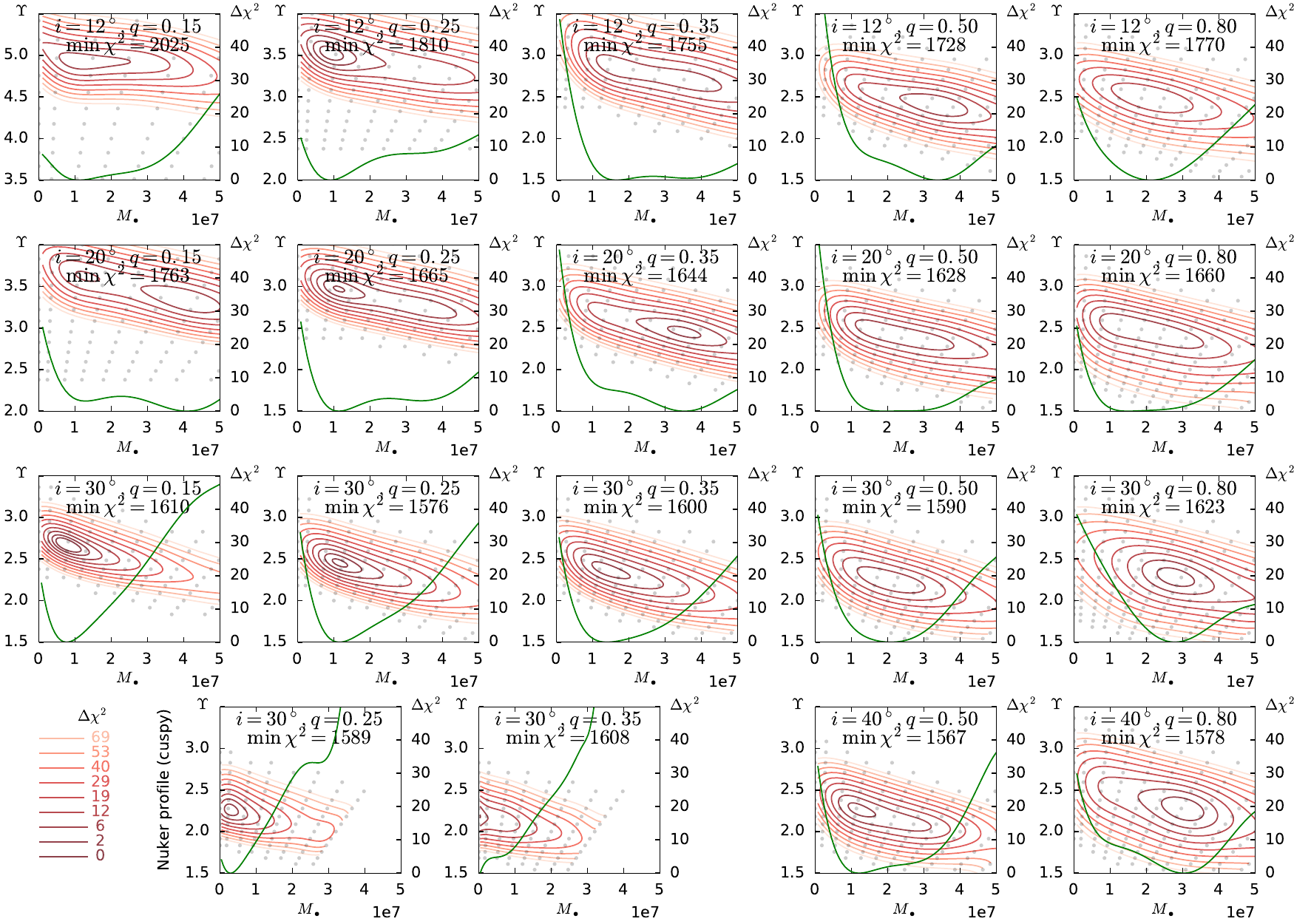}
\caption{Contours of $\Delta\chi^2 \equiv \chi^2-\chi^2_\mathrm{min}$ as a function of two model parameters (black hole mass \mbh and mass-to-light ratio $\Upsilon$) in each panel, for a series of models constrained only by the NIFS kinematics. The contours are placed at $\Delta\chi^2=2.3,6.2,11.8,\dots$, equivalent to $1\sigma,2\sigma,3\sigma$ confidence intervals for two degrees of freedom. The inclination varies from top to bottom row as $i=12^\circ,\,20^\circ,\,30^\circ,\,40^\circ$, and the flattening is $q=0.15,\,0.25,\,0.35,\,0.5,\,0.8$ from left to right column, except the two leftmost panels in the bottom row, which show models with the cuspy Nuker density profile with $i=30^\circ,\,q=0.25,0.35$ (the remaining models use our fiducial Zhao profile). The marginalized 1d intervals of $\Delta\chi^2$ as a function of \mbh alone are shown by green lines in each panel. Gray dots show the actual values of \mbh and $\Upsilon$ for the grid of models.
} \label{fig:chi2_nifs}
\end{figure*}

\begin{figure*}
\includegraphics{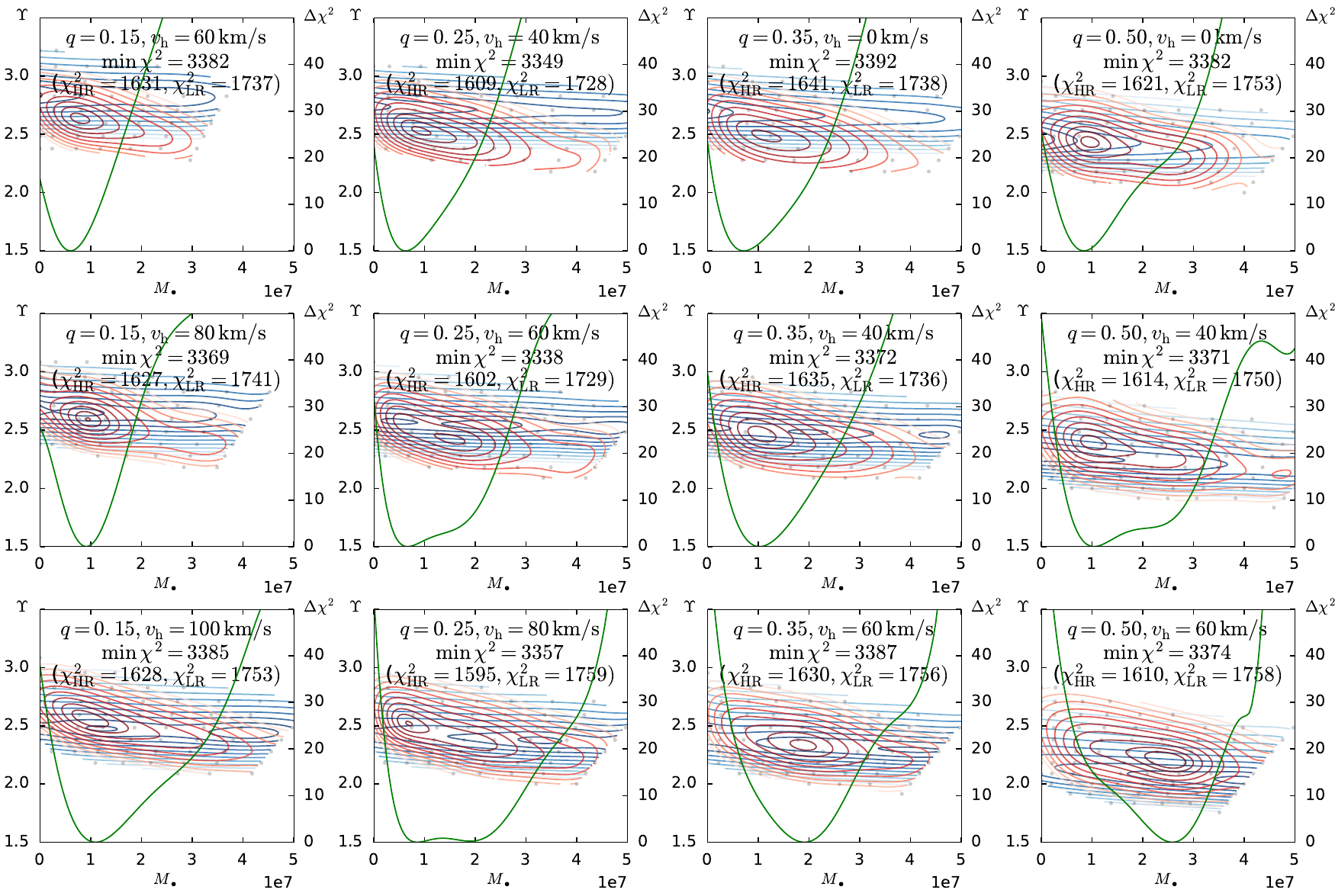}
\caption{Contours of $\Delta\chi^2 \equiv \chi^2-\chi^2_\mathrm{min}$ as a function of two model parameters (black hole mass \mbh and mass-to-light ratio $\Upsilon$) in each panel, for a series of models constrained simultaneously by NIFS and SAURON kinematics. In this case, we plot separately the contours corresponging to both kinematic datasets: NIFS in red (as in Figure~\ref{fig:chi2_nifs}), SAURON in blue; the spacing between contours is the same as in the previous plot. All models in this series use the Zhao density profile and have the inclination angle $i=30^\circ$, while the flattening varies from left to right as $q=0.15,\,0.25,\,0.35,\,0.5$, and the normalization of the dark matter halo increases from top to bottom, parametrized by $v_\mathrm{h}$ (the ranges differ between columns). The marginalized 1d intervals of the total $\Delta\chi^2$ as a function of \mbh alone are shown by green lines in each panel.
} \label{fig:chi2_both}
\end{figure*}

\begin{figure}
\includegraphics{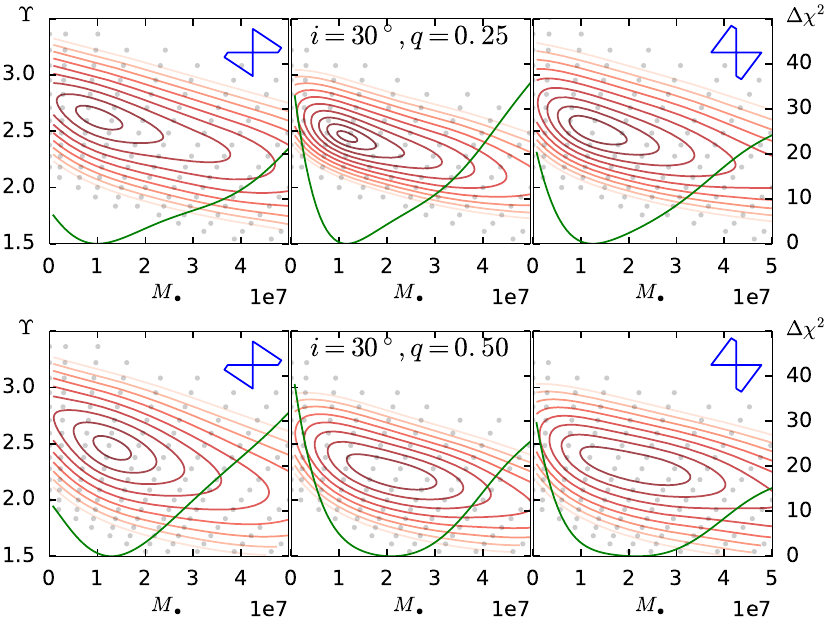}
\caption{Contours of $\Delta\chi^2 \equiv \chi^2-\chi^2_\mathrm{min}$ for variants of models constrained either by the full NIFS datacube (central panels) or by two independent pairs of quadrants separately. The point-symmetric kinematic maps consist of two essentially independent pairs of opposing quadrants (illustrated by blue boundary polygons in the left and right columns).
The central panels are identical to the 2nd and 4th panels in the 3rd row of Figure~\ref{fig:chi2_nifs} (inclination $i=30^\circ$, flattening $q=0.25$ and $q=0.5$); these models are fitted to both variants of kinematic maps simultaneously. The left and right columns instead are fitted to only one of these variants. Despite some differences in the precise location of minima, the shapes of the $\chi^2$ contours are qualitatively very similar between the three variants (although the contours are necessarily tighter in the central column, which has twice as many constraints), thus we conclude that the results are robust to small variations in the observed kinematics.
}\label{fig:chi2_nifs_half}
\end{figure}

Given the relatively large number of free parameters ($i$, $q$, $v_\mathrm{h}$, \mbh, and $\Upsilon$), we do not attempt to cover this parameter space exhaustively, but use a multi-stage strategy.

It is clear that \mbh is only constrained by the NIFS kinematics, since the SAURON data has lower spatial resolution, and moreover, we excise the central few arcsec of SAURON dataset because of AGN contamination. At the same time, the small spatial coverage of NIFS makes it insensitive to the dark matter halo normalization $v_\mathrm{h}$, which is merely a nuisance parameter in the present context. Since our primary goal is to determine \mbh, we first consider a series of models fitted to NIFS kinematics alone and ignore the dark halo.

Figure~\ref{fig:chi2_nifs} shows a four-dimensional grid of models in the parameter space of inclination $i$ (increasing from top to bottom rows), intrinsic axis ratio $q$ (increasing from left to right columns), and in each panel, black hole mass \mbh (abscissa) and mass-to-light ratio $\Upsilon$ (ordinate). 
The two leftmost panels in the lowest row show examples of models that use the cuspy Nuker density profile, and the best-fit \mbh remains near zero for all such models regardless of $i$ and $q$.
The radius of influence $r_\mathrm{infl} \equiv G\,\Mbh/\sigma^2$ for a $2\times10^7\,M_\odot$ black hole and $\sigma\simeq 100$~km\,s$^{-1}$ is $\sim 0\farcs1$, at the limit of resolution of both kinematic and photometric data. Naturally, the stellar mass within this radius is comparable to \mbh, but also varies by a factor of a few between the models with cuspy ($\sim4\times10^7\,M_\odot$) or cored ($\sim1.5\times10^7\,M_\odot$) profiles. We therefore conclude that a cuspy Nuker profile has too high stellar mass in the innermost region, which obviates the need for a black hole. As this is clearly in contradiction with observational evidence for a black hole as demonstrated by AGN activity, in the rest of the paper we consider only the models with the cored Zhao profile; we obtained very similar results with the cored MGE profile from \citet{2014ApJ...791...37O}.

The lowest values of $\chi^2$ differ between panels, but the contours of $\Delta\chi^2 \equiv \chi^2(\Mbh, \Upsilon) - \chi^2_\mathrm{min}$ look similar in all panels, showing a ``tilted valley'' of acceptable models: higher \mbh values correspond to lower $\Upsilon$, such that the total gravitating mass within the region $\lesssim 0.5-0\farcs6$ is approximately constant.
In most panels, the marginalized 1d profiles of $\Delta\chi^2$ as a function of \mbh, shown by green curves, have large, shallow minima in the range $0.5\times10^7\,M_\odot \lesssim \Mbh \lesssim (3-4)\times10^7\,M_\odot$.

Figure~\ref{fig:chi2_both} shows a series of models constrained by both NIFS and SAURON kinematics, with each panel plotting the contours of $\Delta\chi^2$ as functions of \mbh and $\Upsilon$ for a given choice of other parameters. We fix the inclination to $i=30^\circ$ (the overall trends are similar for other values of $i$), and consider different choices of flattening $q$ (thickness increases from left to right) and dark matter halo normalization $v_\mathrm{h}$ (increases from top to bottom).
We plot the contributions of NIFS and SAURON datasets to the total $\chi^2$ of each model separately by red and blue contours.
It is clear that the NIFS contours are very similar to the ones in the third row of Figure~\ref{fig:chi2_nifs}, independently of $v_\mathrm{h}$, and they constrain a certain linear combination of $\Upsilon$ and \mbh, as discussed earlier. By contrast, the SAURON contours are insensitive to \mbh, but constrain a combination of $\Upsilon$ and $v_\mathrm{h}$.
One could reasonably guess that a ``universally acceptable'' model must have compatible best-fit $\Upsilon$ from both datasets, which indeed minimizes the overall $\chi^2$. By adjusting the halo normalization $v_\mathrm{h}$, we can shift the location of best-fit SAURON $\Upsilon$ up or down, and it is natural to select such a value that maximizes the overlap with best-fit NIFS $\Upsilon$. Therefore, the dark matter halo normalization is not really an independent free parameter in our case, but rather should be determined from the consistency between the two datasets.

Comparing the series of models with different flattening $q$, we see that more disky models (smaller $q$) have higher best-fit $\Upsilon$ (especially for low inclination) and correspondingly lower \mbh. This can be understood as follows: since we observe the galaxy at a nearly face-on orientation, the line-of-sight velocity dispersion $\sigma$ is determined by how far the stars travel in the vertical direction (i.e., thickness) and how large is the restoring force (i.e., the mass density). Indeed, for a thin isothermal disk with a surface mass density $\Sigma$ and scale height $h$, the vertical velocity dispersion is $\sigma = \sqrt{2\pi\,G\,\Sigma\,h}$ \citep[problem 4.21]{2008gady.book.....B}. Therefore, when decreasing $q$ towards more disky models and hence decreasing $h$, we must simultaneously increase $\Upsilon$ and hence $\Sigma$ to keep the velocity dispersion at the observed value. On the other hand, when adding a dark matter halo, we increase the gravitating mass and hence reduce the stellar $\Upsilon$, but only in the outer parts (i.e., in the SAURON dataset), where stars are not overwhelmingly dominating the total potential. This suggests that the flattening $q$ is largely degenerate with $\Upsilon$ and $v_\mathrm{h}$, but nevertheless we find that the overall $\chi^2$ is lower for values of $q$ in the range $0.2-0.4$.

Finally, the assumed inclination does have a significant impact on the model properties.
The mean rotational velocity of stars in the equatorial plane $\overline{v_\phi}(R)$ cannot exceed the circular velocity $v_\mathrm{circ}(R) \equiv \sqrt{R\,\partial\Phi/\partial R}$. In the case of a relatively cold disk, the difference (called asymmetric drift) is $v_\mathrm{circ}-\overline{v_\phi} \propto \sigma_R^2$ \citep[][equation 4.228]{2008gady.book.....B}.
The projected rotational velocity is $\sim \overline{v_\phi}\,\sin i$, but for small inclination angles, the line-of-sight velocity dispersion is nearly independent of $i$, being determined by the mass density and thickness of the galaxy. Thus the observed line-of-sight velocity gradient sets the lower limit on the inclination angle $i_\mathrm{min}$, for which stars need to move on nearly circular (``cold'') orbits to produce the given rotational signal. For larger inclinations, stellar orbits have to be ``warmer'' (have higher eccentricity or even rotate in the opposite direction) in order not to exceed the projected rotational velocity.

Comparing different rows of Figure~\ref{fig:chi2_nifs}, we see that models with low inclination generally have higher $\chi^2$ even if the location of the minimum in the \mbh--$\Upsilon$ plane is not strongly varying (except when both the inclination $i$ and the thickness $q$ are so low that the model is forced to have higher $\Upsilon$ to match the line-of-sight velocity dispersion). The tendency of orbit-superposition models to produce better fits at high inclination angles (edge-on orientations), even if the actual orientation is closer to face-on, has been noted in many studies (e.g., Section~5.3 in \citealt{2007MNRAS.381.1672T}, or \citealt{lipka2021}), and can be attributed to a greater flexibility in rearranging the orbits in the case of sub-maximal rotation. However, as discussed in Section~\ref{sec:inclination}, NGC 4151 appears fairly round at large radii, and certainly resembles a disk galaxy seen close to face-on rather than an intrinsically round galaxy seen edge-on. For this reason, we only consider models with $i\le30^\circ$, with an exception of two models with $i=40^\circ$ and $q\ge0.5$. In any case, the range of acceptable values of \mbh does not strongly depend on the inclination, and we take $i=30^\circ$, $q=0.25$ for our fiducial series of models.

As mentioned in Section~\ref{sec:nifs_kinem}, point-symmetrized NIFS kinematic maps provide two independent variants of observational constraints and allow us to test the sensitivity of global model parameters to some asymmetries in the LOSVD. Indeed, the kinematic maps shown in Figure~\ref{fig:nifs} are not symmetric with respect to reflection about the kinematic major axis (exchanging the upper and lower halves of the maps), although they are symmetric (for even moments) or antisymmetric (for odd moments) between the opposite quadrants (upper left and lower right quadrants are identical and represent one variant of data reduction, while upper right and lower left quadrants represent another, independent variant). In most cases we fitted both pairs of quadrants simultaneously, but in Figure~\ref{fig:chi2_nifs_half} we show the results obtained from a series of models constrained by only one of the two variants of kinematic maps. The similar morphology of the $\chi^2$ contours supports the robustness of our results with respect to small variations in the kinematic maps.

\begin{figure*}
\includegraphics{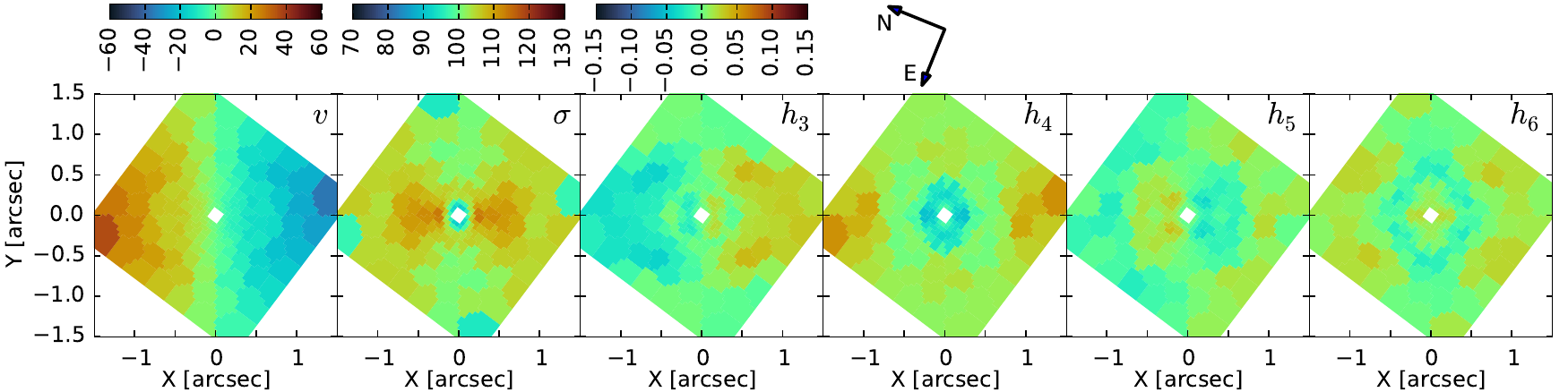}
\caption{Kinematic maps of the central region (the NIFS dataset) for the fiducial model with $i=30^\circ$, $q=0.25$,  $\Mbh=1.3\times10^7\,M_\odot$, and $\Upsilon=2.5$; these maps can be directly compared to the observations shown in Figure~\ref{fig:nifs}.
} 
\label{fig:model_nifs}
\end{figure*}

\begin{figure}
\includegraphics{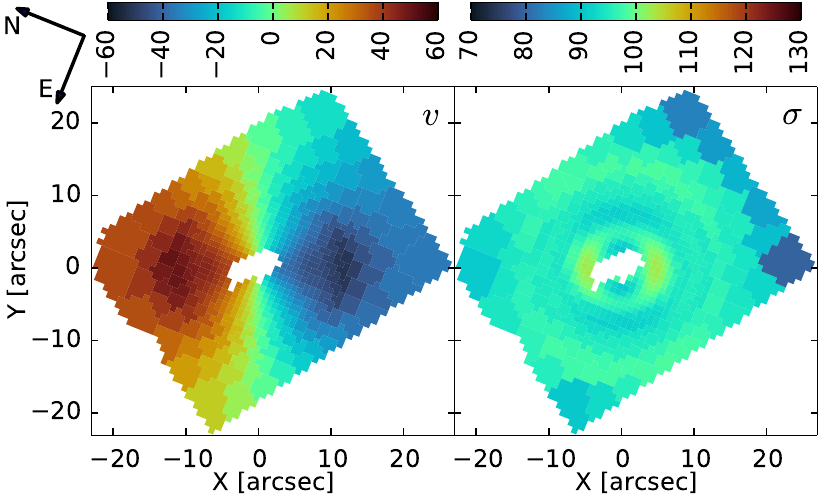}
\caption{Kinematic maps ($v$ and $\sigma$) of the larger region (the SAURON dataset) for the model with $i=30^\circ$, $q=0.25$, $Mbh=1.3\times10^7\,M_\odot$, $\Upsilon=2.5$, and $v_\mathrm{h}=60$\,km\,s$^{-1}$, which can be compared to the observations shown in Figure~\ref{fig:sauron}. The input data did not have higher-order Gauss--Hermite moments, but we did constrain them in the model to have zero mean and associated formal uncertainty of 0.05; the actual values in the models are indeed fairly small and contribute negligibly to the total $\chi^2$.
} \label{fig:model_sauron}
\end{figure}

Figures~\ref{fig:model_nifs} and \ref{fig:model_sauron} show the NIFS and SAURON kinematic maps, respectively, for a fiducial model with $i=30^\circ$, $q=0.25$, $\Mbh=1.3\times10^7\,M_\odot$, and $\Upsilon=2.5$, which has one of the lowest $\chi^2$ values. The maps actually look very similar for other nearby choices of parameters, even though the difference in $\chi^2$ may be rather significant (of order few tens). Variation of the black hole mass most noticeably affects $\sigma$ and $h_4$ maps in the innermost $\sim 0\farcs2$. 

\subsection{Black Hole Mass}

\begin{figure}
\includegraphics{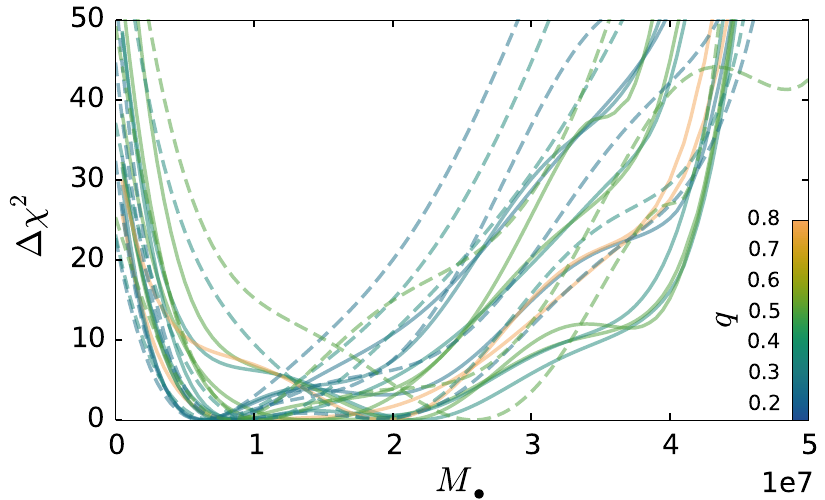}
\caption{Marginalized 1d plots of $\Delta\chi^2$ as a function of \mbh. Each curve shows the difference $\chi^2-\chi^2_\mathrm{min}$ for a series of models with a fixed inclination, flattening (shown by color) and dark matter fraction, where the minimum value varies between series.
Solid lines show NIFS-only models, which generally have broader range of acceptable black hole masses, and dashed lines -- NIFS+SAURON models.
} \label{fig:chi2_marg}
\end{figure}

Having considered the overall trends in the grid of models, we finally return to the main science question of this study -- the measurement of the black hole mass. Summarizing the above discussion, we find that the addition of the SAURON kinematics does not tighten constraints on \mbh, since it also brings another free parameter (dark matter normalization) that is largely degenerate with $\Upsilon$.  The limited spatial extent of the SAURON data does not allow us to detect the spatial gradients of dynamical mass-to-light ratio associated with the gradual increase of the halo contribution with radius and hence to disentangle these two parameters. Since the halo properties are irrelevant for the present study,  we focus primarily on the NIFS-only series of models (Figure~\ref{fig:chi2_nifs}), most of which have similar and rather broad ranges of \mbh masses with small $\Delta\chi^2$. 

It is difficult to establish statistically strict constraints on \mbh for several reasons. First, the discrete nature of orbit-superposition models necessarily implies some noise in the value of $\chi^2_\mathrm{min}$. For each of the panels in that figure, we considered several random realizations of the orbit library, which exhibited variations of $\chi^2_\mathrm{min} \sim \mathcal O(10)$, and the locations of the minima were randomly scattered within regions roughly bounded by $\Delta\chi^2 \lesssim 10$ (see, e.g., bottom row of Figure~2 in \citealt{2020ApJ...889...39V}); in the plot, we show the contours averaged across these realizations.
Second, the value of $\chi^2$ per constraint (reduced $\chi^2$) is $\sim 1.7$, indicating a moderately poor fit (or, more likely, underestimated measurement uncertainties). Third, there is little systematic study of statistical foundations of the Schwarzschild method, in particular, regarding the rigorous choice of confidence intervals on model parameters in terms of $\Delta\chi^2$. This choice cannot be guided by standard considerations applicable to the normally distributed errors (e.g., $\Delta\chi^2=1,4,\dots$ for $1\sigma$, $2\sigma$, etc.), since this does not take into account the highly non-parametric nature of the method: for each choice of ``real'' model parameters (\mbh, $\Upsilon$, etc.), we consider the $\chi^2$ value produced by only one of many possible combinations of orbit weights (``hidden'' parameters), instead of marginalizing over them (see \citealt{2006MNRAS.373..425M} for a discussion). 

Pending a rigorous statistical analysis, we adopt a simple qualitative prescription guided by our experiments with a large suite of models. Namely, we take the 1d $\Delta\chi^2$ profile, marginalized over $\Upsilon$ in each panel, and plot these profiles for several choices of $i$, $q$, and in the case of NIFS+SAURON models, $v_\mathrm{halo}$, to examine the overall range of acceptable \mbh values. Figure~\ref{fig:chi2_marg} shows $\sim20$ such profiles, whose minima span the range $0.5\times10^7\,M_\odot \lesssim \Mbh \lesssim 2\times10^7\,M_\odot$; given the inherent noise in $\chi^2\gtrsim \mathcal O(10)$, we extend the range of acceptable values to $0.25\times10^7\,M_\odot \lesssim \Mbh \lesssim 3\times10^7\,M_\odot$. The corresponding mass-to-light ratios lie in the range $\Upsilon \simeq 2.5\pm 0.3$, consistent with the photometric estimates (Section~\ref{sec:mass_to_light}).

\section{Discussion} \label{sec:diss}

\subsection{Comparison with Previous Results}

\begin{deluxetable}{lcc}
\renewcommand{\arraystretch}{2}
\tablecolumns{3}
\tablewidth{0pt}
\tablecaption{Measurements of \mbh\ for NGC\,4151}
\tablehead{
\colhead{Method} & 
\colhead{\mbh\ ($10^{7}$ M$_{\odot}$)} & 
\colhead{Reference}
}
\startdata
%
SD Modeling & 0.25--3.0 & This work \\
SD Modeling & $4.27^{+1.31}_{-1.31}$ & \cite{2014ApJ...791...37O}\\
GD Modeling & $3.6^{+0.9}_{-2.6}$ & \cite{2008ApJS..174...31H}\\
H$\beta$ RM & $3.57^{+0.45}_{-0.37}$ & \cite{2006ApJ...651..775B}\\
H$\beta$ RM & $2.06^{+0.05}_{-0.05}$ & \cite{2018ApJ...866..133D} 
\label{tab:mbh}
\enddata

\tablecomments{All masses from dynamical modeling have been adjusted to an assumed galaxy distance of 15.8\,Mpc.}
\end{deluxetable}

The results presented in the previous section generally agree with those of  \citet{2014ApJ...791...37O}, which is unsurprising given the large error bars in both studies. \citet{2014ApJ...791...37O} found $\Mbh=(3.76 \pm 1.15)\times10^7\,M_\odot$ but assumed a distance of $D=13.9$\,Mpc.  When adjusted for the recently published Cepheid distance to NGC\,4151 of $D=15.8\pm0.4$\,Mpc \citep{yuan2020}, their best-fit mass becomes $\Mbh=(4.27 \pm 1.31)\times10^7\,M_\odot$. Our range of \mbh values is somewhat lower, which may be attributed to a somewhat lower velocity dispersion $\sigma$ in the central parts inferred from our reanalysis of the observational data. Their mass-to-light ratio $\Upsilon_H=0.34 \pm 0.03$ is quoted for a different frequency band, and is generally consistent with our inferred $\Upsilon_V=2.5 \pm 0.3$, as discussed in Section~\ref{sec:mass_to_light}. 

We also list in Table \ref{tab:mbh} all other \mbh determinations for NGC\,4151 from other direct methods. As we did for the \citet{2014ApJ...791...37O} SD mass, we have adjusted the GD modeling mass to account for our adopted  distance of 15.8\,Mpc. We have also adjusted the RM masses so they have the same adopted value of $\langle{f}\rangle = 4.3$  \citep{2013ApJ...773...90G}.  The SD mass range we report here generally agrees with the two RM masses from \cite{2006ApJ...651..775B} and \cite{2018ApJ...866..133D} and the H$_{2}$ GD modeling mass of \cite{2008ApJS..174...31H}. 

\subsection{Limitations of the Dynamical Models}

Despite the improvements in the data reduction and analysis process, our uncertainties on \mbh are much larger than in previous studies of the same galaxy, primarily  due to the fact that we have explored a wide range of parameters but have only reported results for marginalizing over $\Upsilon$.
We acknowledge that the intrinsic shape and orientation of the galaxy are not well constrained, and instead of considering only one formally best-fit choice of these nuisance parameters, we accepted the \mbh values from a broad range of models. However, Figures~\ref{fig:chi2_nifs}, \ref{fig:chi2_both} demonstrate that even for a given choice of galactic geometry, there is a relatively large uncertainty on \mbh made possible by suitable rearrangement of orbits in our extensive orbit libraries. This confirms the expectations stemming from the analysis of mock data in Section~3.2 of \citet{2020ApJ...889...39V}, but stands at odds with more optimistic conclusions about the precision of \mbh measurements reached by some other studies, e.g., \citet{2021MNRAS.500.1437N}. As these experiments did not use the same setup, it is clear that a more thorough investigation of this question is needed in the future.

The rather weak constraints on \mbh that we obtain in this study may seem overly pessimistic, but we recall that this galaxy is a rather difficult case from the observational perspective, since the AGN strongly dominates the light profile and limits the precision of kinematic measurements at small radii. 
For our estimated mass range $0.25\times10^7\,M_\odot \lesssim \Mbh \lesssim 3\times10^7\,M_\odot$, and our measured central value of $\sigma\sim 100$km~s$^{-1}$ in the inner few pixels, the sphere-of-influence of the black hole is $1.1-11.2$pc or 0$\farcs015-0\farcs$15. This implies that the radius of influence of the black hole is barely resolved by the innermost few pixels in the NIFS dataset. Moreover, the stellar luminosity profile in the same central region is also very difficult to constrain due to overwhelming AGN brightness, and as we have seen, for a plausible cuspy stellar profile consistent with observations, dynamical models prefer no black hole, which is clearly counterfactual. These considerations highlight the importance of spatially resolving the sphere of influence both photometrically and kinematically, 
especially in late type galaxies where black hole masses and stellar velocity dispersions are typically smaller than in elliptical galaxies.

Previous authors \citep[e.g.][]{Gultekin_2009, Batcheldor_2010, Gultekin_2011} have focused on the estimated parameters of the \mbh-$\sigma$ relation when \mbh measurements that do not resolve the black hole sphere-of-influence are included or excluded from the sample. \citet{Gultekin_2011} found, for a sample of \mbh in early type galaxies with central velocity dispersions with median $\sigma \gtrsim 260$km~s$^{-1}$, that excluding measurements which do not resolve the radius-of-influence of the black hole biases the estimated \mbh-$\sigma$ relation. They concluded that instead of being excluded these measurements should be included as upper limits or with large error bars. For the elliptical galaxies and massive bulges in these studies, it is argued based on empirical evidence from repeated measurements, that resolving the radius-of-influence only helps to reduce the errors on the black hole mass estimate.

Using mock long-slit kinematic data \citet{2004ApJ...602...66V} showed that if the radius-of-influence of a black hole is not resolved, nuclear kinematical data can be fitted without a black hole. As far as we are aware the work presented in this paper is the first modeling study in a late type galaxy that clearly demonstrates the drawbacks of not resolving the influence radius on the stellar dynamical \mbh measurement. Despite the fact that the AGN makes it difficult to resolve the influence radius of the black hole, its very presence necessitates a black hole. Yet we saw that because the central luminosity profile is not resolved below 0\farcs1, the assumption of a cuspy Nuker density profile (which is generally considered reasonable for a late type galaxy) would require no black hole to fit the kinematics. Therefore this study clearly illustrates the need for both photometric and kinematic data that resolve the influence radii of supermassive black holes, data that will become more readily available with JWST and the ELTs.

\subsection{Future Prospects}

Efforts to model the RM observations presented by \cite{2018ApJ...866..133D} are currently underway and will remove the dependence on $\langle{f}\rangle$ and provide a fully-independent RM mass for comparison with the SD and GD-based masses.  Such analyses rely on  strong AGN variability during the monitoring campaign, as well as careful management of all noise sources and the observing cadence, and have only been possible for a handful of objects thus far. The observations presented by \cite{2018ApJ...866..133D} are of similar quality to previously successful RM modeling attempts \citep{2014MNRAS.445.3073P,2017ApJ...849..146G,williams18}, and should therefore be sufficient for an accurate mass constraint.

Furthermore, NGC\,4151 is the target of an Early Release Science program with JWST (ERS 1364, PI Bentz).  Observations with the NIRSpec IFU will probe the nuclear stellar dynamics and will be directly compared with the observations and analysis presented in this work.  NIRSpec is expected to provide some crucial advantages over AO-assisted ground-based observations with its stable and diffraction-limited PSF and significantly lower backgrounds.  However, NIFS provides a higher spectral resolution, which may allow for more accurate measurements of the higher-order moments of the stellar absorption profiles.  We will thus revisit the topic of the black hole mass in NGC\,4151 from stellar dynamical modeling in the near future, once JWST has successfully launched.

\section{Summary} \label{sec:sum}

We have presented a full reanalysis of the black hole mass derived from nuclear stellar dynamics in NGC\,4151, beginning with the raw data cubes observed with Gemini NIFS and Altair adaptive optics.  We implemented several improvements to the data reduction, including modifications to the NIFS pipeline to allow the variance information to be carried through the full reduction process to the final combined cubes.  We also improved the spectral measurements derived from the final combined  data cubes, preserving the spatial resolution near the central AGN through the use of adaptive binning, and jointly fitting the spectra for point-symmetrized  bins rather than bisymmetrizing the Gauss-Hermite terms afterwards. 
With the new orbit-superposition code \textsc{Forstand}, we conducted a thorough exploration of the parameter space within the dynamical models, investigating the use of a dark matter halo and including models with a range of \mbh, $\Upsilon$, galaxy inclination $i$, and bulge flattening $q$.  We have also adopted the first measurement of an accurate and precise distance to NGC\,4151 based on observations of Cepheid stars.  We find the black hole mass lies within the range of  $0.25\times10^7\,M_\odot \lesssim \Mbh \lesssim 3\times10^7\,M_\odot$, which is in general agreement with results from reverberation mapping and gas dynamical modeling.  Future dynamical modeling of reverberation data as well as IFU observations with JWST will further constrain the mass of the central supermassive black hole in NGC\,4151.

\acknowledgements

We thank the anonymous referee for comments that improved the presentation of this work.  CAR and MCB gratefully acknowledge support from the National Science Foundation through CAREER grant AST-1253702 and AAG grant AST-2009230. 
CAO acknowledges support from the Australian Research Council through Discovery Project DP190100252. MV gratefully acknowledges support from the National Science Foundation through AAG grants AST-1515001 and AST-2009122, and the Space Telescope Science institute through HST-AR-13890.001 and JWST-ERS-01364.002-A. EV acknowledges support from STFC via the Consolidated grant to the Institute of Astronomy. We thank Eric Emsellem for providing SAURON IFU observations of NGC 4151.

This work is  based on observations acquired through the Gemini Science Archive and processed using the Gemini IRAF package, which is managed by the Association of Universities for Research in Astronomy (AURA) under a cooperative agreement with the National Science Foundation on behalf of the Gemini Observatory partnership: the National Science Foundation (United States), National Research Council (Canada), Agencia Nacional de Investigaci\'{o}n y Desarrollo (Chile), Ministerio de Ciencia, Tecnolog\'{i}a e Innovaci\'{o}n (Argentina), Minist\'{e}rio da Ci\^{e}ncia, Tecnologia, Inova\c{c}\~{o}es e Comunica\c{c}\~{o}es (Brazil), and Korea Astronomy and Space Science Institute (Republic of Korea). 
This work was enabled by observations made from the Gemini North telescope, located within the Maunakea Science Reserve and adjacent to the summit of Maunakea. We are grateful for the privilege of studying the Universe with observations that were acquired from a place that is unique in both its astronomical quality and its cultural significance.
The SAURON observations were obtained at the William Herschel Telescope, operated by the Isaac Newton Group in the Spanish Observatorio del Roque de los Muchachos of the Instituto de Astrofisica de Canarias.



\bibliographystyle{apj} 

\end{document}